\documentclass[11pt,a4]{emulateapj}
\slugcomment{{\sc Accepted to AJ:} April 15, 2008} 
\usepackage{amsmath}
\usepackage{lscape}

\shortauthors{Fuentes \& Holman 2008}
\shorttitle{A Subaru Archival Search for Faint TNOs}

\def\Eq#1{\rm Eq.~\ref{#1}}
\def\Fig#1{\rm Fig.~\ref{#1}}
\def\fig#1{\rm Figure~\ref{#1}}
\def\tab#1{\rm Table~\ref{#1}}
\def\km{~\rm km}

\def\au{~\rm AU}

\def\deg{~\rm deg}
\def\sqdeg{~\rm deg^2}
\def\aph{~\rm ''/hr}
\def\arcsec{~\rm ''}
\begin{document}

\bibliographystyle{apj}

\title{A Subaru Archival Search for Faint Trans-Neptunian Objects\footnotemark[1]}
\author{Cesar I.\,Fuentes\altaffilmark{2}, Matthew J.\,Holman\altaffilmark{2} }
\footnotetext[1]{Based on data collected at Subaru Telescope, which is operated by the National Astronomical Observatory of Japan.}
\altaffiltext{2}{Harvard-Smithsonian Center for Astrophysics, 60 Garden Street, Cambridge, MA 02138; cfuentes@cfa.harvard.edu}

\begin{abstract}
  We present the results of a survey for trans-neptunian objects
  (TNOs) based on Subaru archival images, originally collected by
  \citet{Sheppard.2005} as part of a search for irregular satellites
  of Uranus. The survey region covers $2.8\sqdeg$, centered on Uranus
  and observed near opposition on two adjacent nights.  Our survey
  reaches half its maximum detection efficiency at
  $R$=$25.69\pm0.01$. The objects detected correspond to 82~TNOs,
  five~Centaurs, and five~irregular satellites.  We model the
  cumulative number of TNOs brighter than a given apparent magnitude
  with both a single and double power law. The best fit single power
  law, with one object per square degree at magnitude
  $R_0$=$22.6_{-0.4}^{+0.3}$ and a slope of
  $\alpha$=$0.51_{-0.6}^{+0.5}$, is inconsistent with the results of
  similar searches with shallower limiting magnitudes. The best fit
  double power law, with a bright-end slope
  $\alpha_1$=$0.7_{-0.1}^{+0.2}$, a faint-end slope $\alpha
  _2$=$0.3_{-0.2}^{+0.2}$, a differential number density at $R=23$
  $\sigma_{23}$=$2.0_{-0.5}^{+0.5}$ and a magnitude break in the slope
  at $R_{eq}$=$24.3_{-0.1}^{+0.8}$, is more likely than the single
  power law by a Bayes factor of $\sim$26. This is the first survey
  with sufficient depth and areal coverage to identify the magnitude
  at which the break occurs without relying on the results of other
  surveys.

  We estimate barycentric distances for the 73 objects that have 24~hr
  arcs; only two have heliocentric distances as large as
  $\sim$$50\au$. We combine the distribution of observed distances
  with the size distribution that corresponds to a double power law
  luminosity function to set a tight constraint on the existence of a
  distant TNO population.  We can exclude such a population at
  60$\au$, with 95\% confidence, assuming it has the same size
  distribution and albedo as the observed TNOs, if it exceeds 8\% of
  mass of the observed TNOs.

\end{abstract}

\section{Introduction}\label{sec:intro}
The remnants of the protoplanetary disk, now in the form of
trans-neptunian objects (TNOs), offer a unique way to study the
evolution of the solar system. The TNO size distribution is defined by
its initial properties, collisional history, and the formation and
evolution of the giant planets \citep{Kenyon.2004, Pan.2005,
Kenyon.2007}. The orbital dynamics of the TNOs is largely governed by
interactions with Neptune, and the radial distribution of TNOs also
depends on the giant planets' evolution (see \citealt{Morbidelli.2007}
for a review).  It has been suggested that the radial extent of TNOs
was truncated by a close passage of a star during the early stages of
the Solar System formation \citep{Brunini.1996,Ida.2000,
Kobayashi.2001, Kenyon.2004}.

A number of large-scale investigations that will significantly advance
our understanding of the outer solar system are currently being
designed, tested, and executed. Pan-STARRS~\citep{Jewitt.2003}, given
its coverage of the sky and time baseline, promises an accurate
determination of the statistical properties of the trans-neptunian
region.  LSST~\citep{Tyson.2001} and SkyMapper~\citep{Keller.2007}
will extend the surveyed sky to the southern hemisphere. The New
Horizons (NH) mission will give unprecedented views of the
trans-neptunian space by approaching Pluto and other TNOs in mid-2015.
Nevertheless, there remain important questions that will not be
answered by these studies. These large synoptic surveys will
necessarily have a shallow limiting magnitude. Deep surveys like this
will continue to be the only window into the smallest and farthest
objects in the Solar System. The answers to these questions can
influence how these projects are carried out and how their resulting
data are interpreted. The distribution of faint objects will matter
when large surveys choose fields to be covered more deeply. The TNO
size distribution and radial extent of TNOs are among the questions
that will rely on pencil-beam surveys to be answered.

Since the discovery of 1992~QB$_1$~\citep{Jewitt.1992} a number of
wide-field surveys for TNOs have been completed \citep{Jewitt.1998,
Chiang.1999, Larsen.2001, Trujillo.2001, Trujillo.2001a,
Millis.2002,Trujillo.2003,Elliot.2005,Larsen.2007}. In addition to
determining much of the dynamical structure of the trans-neptunian
region and identifying large, bright TNOs that are amenable to follow
up observations, these surveys constrain the bright end
($R$$\lesssim$24) of the cumulative luminosity function of TNOs, the
number of objects per square degree brighter than a given
magnitude. This quantity has consistently been measured to be a power
law of the form
\begin{equation}
\Sigma(R) = 10^{\alpha (R-R_0)}  ,
\end{equation}
where $R_0$$\sim$23 is the magnitude at which one expects 1 object per
square degree and $\alpha$ is the slope of the distribution.

Ground based efforts have also focused on detecting fainter TNOs with
deeper imaging of narrow areas of the sky. Many have been successfully
conducted in recent years \citep{Gladman.1998, Gladman.2001,
Allen.2001, Allen.2002, Petit.2004, Petit.2006, Fraser.2008}. These
surveys have been concentrated near the ecliptic plane and reach
limiting magnitudes as faint as $R$$\sim$26. These surveys also find
that the cumulative surface density of TNOs is consistent with a
single power law.  

However, the deepest search to date, using the {\it
Hubble Space Telescope} with the {\it Advanced Camera for Surveys} and
reaching a 50\% detection efficiency at magnitude $R$=28.5, found 25
times fewer objects than expected from extrapolating the brighter
($R$$\leq$25) distribution~\citep{Bernstein.2004}. Their result
indicates there is a break in the cumulative surface density of
objects near $R$$\sim$25. 

\citet{Bernstein.2004} necessarily relied upon the results of other
surveys to assess the deviation of the cumulative density of objects
from a single power law over a range of magnitudes. However, it is
difficult to combine the results of different surveys to obtain a
well-calibrated sample of the trans-neptunian population.

Dynamical biases in latitude and longitude, change the local density
of objects and the relative abundances of excited and classical
objects depending on the direction in which a survey is
conducted. This can be seen in the variety of results found in the
literature; a nice re-analysis and summary of some surveys is
presented by \citealt{Fraser.2008}. Different surveys sample a variety
of ecliptic latitudes and longitudes, use various analysis methods, or
vary in observing conditions. It is necessary to determine and correct
for the effects of these differences to characterize the physical and
dynamical properties of the TNO population. For bright TNOs, large
synoptic surveys will determine many of the biases in the observations
as well as in the population itself. However, for fainter TNOs, the
simplest way to overcome these difficulties is to observe a single
region of the sky.

All these surveys use the ``digital tracking'' method
\citep{Gladman.1997, Gladman.1998, Gladman.2001, Allen.2001,
Allen.2002}, by which a series of consecutive short exposures are
digitally shifted and coadded to match the apparent motion of real
objects. That method has proven very useful in improving the
sensitivity of these ground and space based observations. However,
this method relies on how fine the grid of velocities sampled is, the
good quality of a template image to subtract from each exposure, extra
processing of the images and a trained operator to filter false
positive detections due to saturated stars or other artifacts. Our
results were obtained by linking detections in three different images
described in \S\ref{sec:mod}. Our method's data reduction is more
direct, requires less human interaction and is easier to
photometrically calibrate.

The radial extent of the classical TNO population is not known with
certainty. Although there is evidence for a sudden decrease in density
at $r\sim50\au$ \citep{Trujillo.2001, Gladman.2001}, the existence of
a second farther population near the ecliptic is difficult to rule
out, due to the bias against detection of more distant, fainter
objects. We are slightly more sensible to distant, slower moving
objects. Since we do not rely on the construction of a template field,
usually made with data taken on the same night, that increases the
noise and would subtract signal from very slow movers.

The objectives of this work are to better constrain the expected break
in the TNO luminosity function using a single survey and to better
understand the lack of detections at large heliocentric distances. In
the next section we describe the data and the processing of images.  In
\S~\ref{sec:mod} we present our moving object detection algorithm.  We
discuss the control population and detection efficiency of our method in
\S~\ref{sec:effic}. In the final two section we present the results of
our survey and disucss their implications for the size and distance
distribution of the TNO population.

\section{Data}\label{sec:data}
The observations considered in this project were taken on UT 2003
August 29 and 30 with
Suprime-Cam~\citep{Miyazaki.2002} mounted on the Subaru
telescope. Suprime-Cam is a mosaic camera with 10 CCDs, each with 2048
$\times$ 4096 pixels. Each mosaic image has a field of view of $34'
\times 27'$. We used SMOKA, the electronic archive of the Subaru
Telescope~\citep{Ichikawa.2002}, to retrieve observations taken in
August 2003 in the vicinity of Uranus, near opposition. The fields
were originally observed by \citet{Sheppard.2005}. They surveyed a
total of 14 fields, with an areal coverage of $3.57\sqdeg$ over the
course of two nights. All exposures were taken with the ``Cousins R''
red filter, well-matched to the colors of outer solar system objects.

The objective of the original investigation was to discover uranian
irregular satellites. \citet{Sheppard.2005} recovered all previously
known uranian irregular satellites and discovered two new such
satellites. The faintest satellites detected have magnitudes at
R$\sim$25.5 \citep{Sheppard.2005, Kavelaars.2004}. On the first night,
the observers took two or three exposures of $\sim$7 min of each
field, separated by half an hour on the first night. They re-observed
those fields with two exposures during the second night, with the
pointings shifted to maintain the same positions relative to
Uranus. The survey was designed to discover objects during the first
night and to obtain better orbital information using the second
night's data.

We chose this particular data set for the following reasons. The data
set is sensitive to $R \lesssim 25.5$ magnitude objects, in the
magnitude range in which \citet{Bernstein.2004} find the break in the
TNO cumulative function to be. This sensitivity is reached in a single
exposure, avoiding the difficulties associated with combining
different images. There are 11 fields ($2.8\sqdeg$) with 3 exposures
on the first night, permitting a simple search for moving objects. The
fields were observed very close to opposition, allowing a reliable
distance estimate from the rate of motion with only a 24 hr arc. The
sky coverage is large enough to expect the discovery of $\sim$100
TNOs, allowing a significant constraint on the cumulative
luminosity function. Finally, the data were easily obtained from the SMOKA
system, after the 18 month proprietary period.

We performed the usual calibration of the images. For every image, we
performed an overscan correction, trimming, bias frame subtraction,
and flat-field division using standard IRAF\footnote{IRAF is
distributed by the National Optical Astronomy Observatories, which are
operated by the Association of Universities for Research in Astronomy,
Inc., under cooperative agreement with the National Science
Foundation.} routines. Calibration frames taken during these
observations were obtained from SMOKA.

\section{Moving Object Detection}\label{sec:mod}
The apparent motion of outer solar system objects viewed near
opposition is primarily due to the Earth's translation. For objects at
the distance of Uranus the apparent motion can be as large as
$\sim$6$\aph$. For TNOs this rate is typically $\sim$3$\aph$. This
motion, with respect to background stars, is readily detected even in
the short ($\sim$1 hour) time baseline of this dataset.

To find TNOs, Centaurs, and irregular satellites in this data set, we
use a variant of the search algorithm described and implemented by
\citet{Petit.2004}.  This method is similar to that used in other TNO
surveys (for example,
\citealt{Levison.1990,Irwin.1995,Jewitt.1995,Trujillo.2001,Millis.2002}).
The algorithm detects moving objects by comparing the positions of all
point sources in each of three images of a patch of sky taken in the
same night.  Thus, as mentioned earlier, we restricted our search to
the 11 fields for which there were 3 images taken on the first night.
The individual steps in the algorithm are as follows.

First, for each search field we determine an astrometric solution for
the first image of the night.  These astrometric solutions are used
later to guide the insertion of synthetic moving objects. We used the
2MASS point source catalog~\citep{Cutri.2003} as an astrometric
reference. The RMS in the astrometric solution was typically of
$0.2\arcsec$ or lower (close to the catalog's precision). The relative
errors on the astrometric solutions for both nights were comparable to
the tolerance of the search algorithm.

We then register the second and third images with the first image of
the night. This allows for very accurate positioning of stellar-like
objects with respect to each other.  This is done for the individual
CCDs, rather than for the entire mosaic.  The successive CCDs images
are linearly interpolated, automatically, to the first using the
positions of the background stars and routines available in the ISIS
package~\citep{Alard.2000}.  When thse routines failed to converge
(due to numerous bad pixels or saturated stars), we align the images
interactively using routines from IRAF.

At this stage, we insert the population of synthetic objects that will
be used to determine the detection efficiency of the search, as
described in \S~\ref{sec:effic}.

We then use two different algorithms to search for point sources. The
first of these is a wavelet transform source detection routine (see
\citealt{Petit.2004} for a description).  The second is the publicly
available SExtractor package~\citep{Bertin.1996}, which calcuates the
local image background RMS, convolves the image with a user-specified
kernel, and then identifies groups of pixels with values exceeding the
background variation by a given value.  These two approaches have very
different false detection characteristics.  Thus, we consider the
intersection of detections from both routines.  (We use the flux
information given by SExtractor for the photometry described in
\S~\ref{sec:results}).  We use a detection threshold of $2.6
\sigma$. For SExtractor this corresponds to four adjacent pixels with
values that are at least 1.3 times of the local background variation.
At the end of this stage, there are three lists of sources, one list
for each image.  This results in up to $\sim$50,000 detection in each
mosaic image.  Note that the expected motion of trans-neptunian
objects during a single exposure ($\sim$7 minutes) is small compared
to the typical FWHM ($0.7\arcsec$).  Thus, trailing does not
significantly affect the source detection.

In order to identify moving objects among all the point source
detected, we apply a series of filters that eliminate individual
detections, as well as sequences of detections, that are not
consistent with the TNO population.

We first reject all detections that corresponded to stationary
objects, {\it i.e.} stars and galaxies.  For each list of detections,
we eliminate those for which there is a corresponding detection within
$0.05\arcsec$ in at least one of the other two lists.  We deliberately
chose a small threshold in order to not diminish our sensitivity to
very slow moving TNOs.  This stage typically reduces the number of
detections to $\sim$10,000 per field.

The next step is to search for linear motion among the non-stationary
detections.  We identify all groups of three detections in the
successive images that are consistent with straight line motione (within
$15^\circ$ of the ecliptic), with a constant angular rate between 0.5
and $10\aph$.  The parameter space is chosen to include the expected
rate and direction of TNOs. We consider all combinations of detections
in the three different exposures whose fit to a line had an RMS of
$0.3\arcsec$ or less. These criteria are met by $\sim$1,600
combinations per field, nearly all of which are synthetic TNOs (see
\S\ref{sec:effic}).

In the final stage, the search program outputs an image with all the
combinations found, showing a stamp centered on every detection. We
visually inspect these images to accept or reject a given
object. images. This method allows the spurious and acceptable
detections to be rapidly distinguished ($\sim$30 min/field). Typically
$\sim$20 objects are rejected in this stage per field, the majority
being optical artifacts, bad pixels, extended objects or some
combination of the above.

\section{Control Population and Detection Efficiency}\label{sec:effic}
Since our observations are flux-limited it necessary to account for
detection biases when estimating the intrinsic number of TNOs as a
function of magnitude.  We characterize our search using a population
of synthetic TNOs inserted just after the images have been calibrated
and their astrometric solutions determined. The procedure is done for
each mosaic field, rather than CCD by CCD.  This process nicely
accounts for the possible motion across detector boundaries. The same
synthetic populations were used for the second night.

We used the {\it Orbfit} routines \citep{Bernstein.2000} to create a
realistic population of synthetic TNOs. The characteristics of the
population were chosen to span the range of observational properties
expected of the TNOs. The position of an object on the sky at the time
of the first exposure was drawn from a uniform distribution that
encompassed the FOV of the mosaic. Objects were implanted with
distances between $20\au$ and $200\au$, or alternatively $0.7\aph$ to
$5.0\aph$. The proper motion and radial velocity given to the object
are taken from a distribution that encompasses the possible rates
found in the Solar System.  This initial position and velocity vectors
are only accepted if they correspond to a bound orbit.  If so, we use
the {\it Orbfit} routines to calculate the RA and Dec of the object at
the beginning and end times of each exposure.  We translate these sky
positions into locations on the mosaic using the astrometric solution
derived earlier.

For each exposure we compute a model for the PSF. The model is the
average of $\sim$10 bright, isolated stars for every CCD.  Given the
known magnitude of the synthetic TNO and the measured zero point, and
accounting for transparency changes through changes in the flux
measured in the PSF stars, we use IRAF routine to insert PSFs with
this flux at the calculated positions. We inserted objects from 22.5
to 26.5 mag, which spans the magnitudes of the TNO population we
expect to find.  The flux of each object includes photon noise.  We
did not consider variable objects, as this is unlikely to be
significant on $\sim$1~hr time scales.  We include the effect of
trailing by dividing the flux among several PSFs inserted at positions
linearly interpolated between those at the beginning and end of the
exposures.  This process takes into account any background,
transparency, seeing, and focus variations that might affect the
limiting magnitude. Using this PSF model from each image, we implant a
set of $\sim$2,000 objects per field.  This results in a sufficient
number of synthetic objects per CCD to sample the detection efficiency
as a function of position.

\begin{figure}[ht]
\epsscale{1.0} \plotone{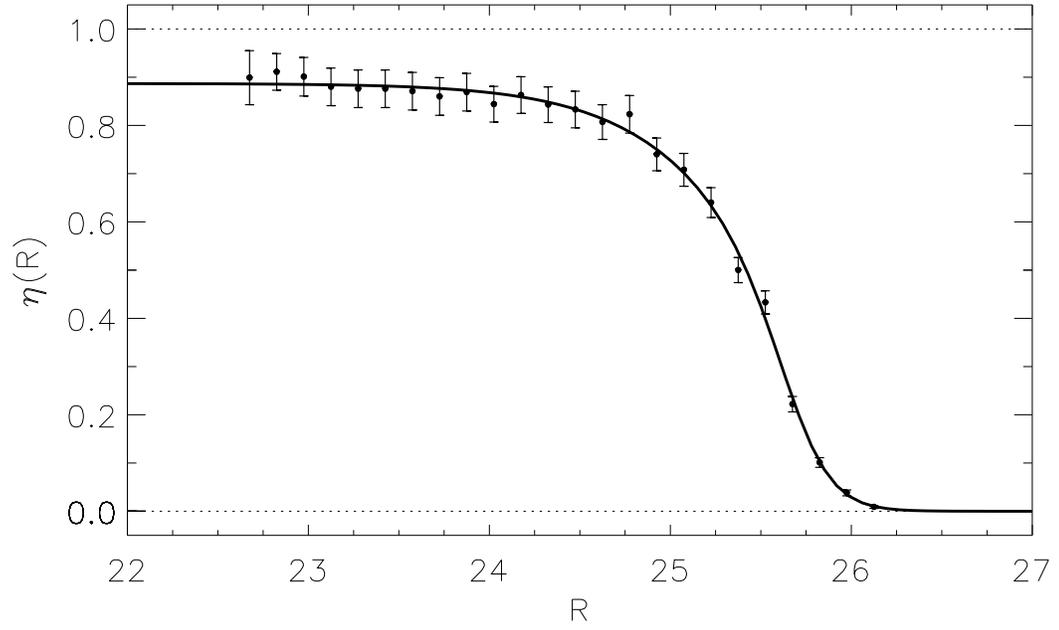}
\caption{\label{fig:effofmag} Detection efficiency as a function of
  magnitude, with an error given by the number of objects implanted
  and found in each bin. The fitted curve corresponds to
  $\Eq{eqn:eff}$, where the best fit values are $A$=$0.88\pm0.01$,
  $R_{50}$=$25.69\pm0.01$, $w_1$=$0.28\pm0.04$ and
  $w_2$=$0.88\pm0.15$. $R_{50}$ corresponds to the magnitude at which
  our method is 50\% as efficient as its maximum detection
  efficiency. }
\end{figure}

Since we are counting objects up to a certain brightness and our model
describes the underlying TNO population, it is essential to estimate
what fraction of the population we detect as a function of
magnitude. In Figure \ref{fig:effofmag} we include a histogram of the
fraction of objects that were recovered in each magnitude bin. We
implanted 25,074 objects in 11 fields, recovering 17,195 of them.

When plotting the cumulative function we used the local efficiency
function, each detection is weighed by the number of objects recovered
in the same field and within the observational magnitude error. The
detection efficiency could vary from field to field. Since all fields
were taken in the vicinity of Uranus, efficiency could depend on
location. However, its statistical effect was negligible on the
efficiency.

For the statistical analysis the effective efficiency function will
need to be integrated. Since it is simpler to integrate analytical
expressions, we used the total efficiency function, that considers all
fields. Following \citet{Petit.2006}, we represent it by 
\begin{equation}\label{eqn:eff}
\eta(R) = 
\frac{A}{4} \left(1-\tanh{\frac{R-R_{50}}{w_1}} \right)
\left(1-\tanh{\frac{R-R_{50}}{w_2}}\right),
\end{equation}
where the best fit values are $A$=$0.88\pm0.01$,
$R_{50}$=$25.69\pm0.01$, $w_1$=$0.28\pm0.04$ and
$w_2$=$0.88\pm0.15$. The errors were obtained with a Markov Chain
Monte Carlo simulation. The parameter $A$ corresponds to the maximum
efficiency, achieved for bright objects. $R_{50}$ corresponds to the
magnitude at which the detection efficiency drops to half the maximum
values.  The parameters $w_1$ and $w_2$ characterize the abruptness of
the decline of the detection efficiency. Figure~\ref{fig:effofmag}
shows the average efficiency function for our data set.

\begin{figure}[ht]
\epsscale{1.0} \plotone{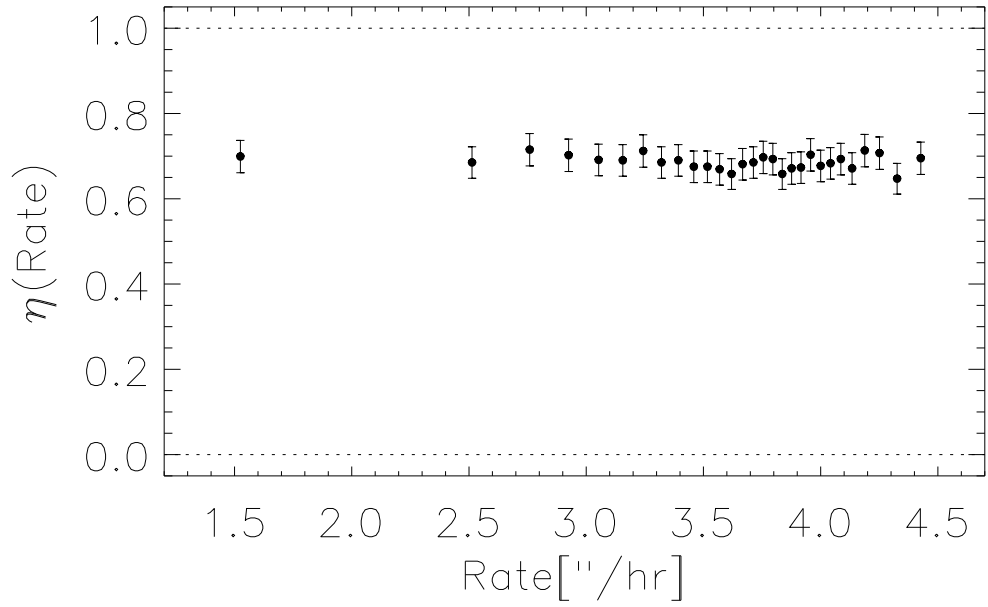}
\caption{\label{fig:effofrate} Histogram of the fraction of objects
  recovered as a function of rate. Bins are chosen to have similar
  numbers of objects.  This demonstrates that our detection efficiency
  does not depend significantly upon the rate of motion.}
\end{figure}

The efficiency could also depend on the rate of motion. We construct a
rate-analog to the magnitude efficiency (See \Fig{fig:effofrate}). The
detection efficiency is nearly indepedent of rate, but our method is
slightly less efficient at larger rates. A faster moving object that
is detected in the first image has a greater chance of falling close
to a background star, or moving outside the field of view, thus the
detection efficiency declines with the rate of motion. The lowest bin
plot in \Fig{fig:effofrate} is $1.5\aph$. Since we implanted objects
to have a population with a constant surface density that bin is not
well sampled. Even though we were able to recover objects planted with
rates as slow as $0.7\aph$ (parallax for objects at $200\au$) we
consider a more conservative limit. The rate at which an object moves
1-FWHM in 45 minutes, the shortest separation between the first and
third exposure, is $0.9\aph$ ($150\au$).

To properly account for detection biases, both real and control
objects must go through exactly the same validation procedure. We did
not unveil the fake object list until all objects were recognized as
moving objects, either real or planted.

\section{Results and Analysis}\label{sec:results}
We found 92 moving objects, five corresponding to known irregular
uranian satellites (those found by \citealt{Kavelaars.2004} and
\citealt{Sheppard.2005}), five to Centaurs, and 82 to TNOs. The
satellites that were missed were blended with stars in one of the
images and hence were not found by our algorithm.

We present our detections in Table \ref{tab:fits}.  For each TNO, we
list its internal designation, its position at the time of the first
exposure (also listed), and its estimated magnitude with uncertainties
(along with an independent estimate of the photometric uncertainty).
We also list the measured sky plane rates of motion of the TNO, two
estimates of the distance to the TNO (one suited for two night's data
and another based only on parallax, both explained later), its orbital
inclination, and separation from Uranus at the time of discovery.

Three standard stars \citep{Landolt.1992} were used to obtain the zero
point and airmass dependence of the photometry. These were PG2213-006C
(V=$15.11\pm0.0045$, V-R=$0.426\pm0.0023$) SA-92-417
(V=$15.92\pm0.0127$, V-R=$0.351\pm0.0151$) and SA-92-347
(V=$15.75\pm0.0255$, V-R=$0.339\pm0.0295$). Since their colors are
similar to those of typical TNOs \citep{Peixinho.2004} we did not
apply a color correction. We checked both nights were photometric and
stable. The possible dependence on seeing (FWHM) was also
investigated, finding it to be unimportant. The correction term was
negligible compared to the airmass correction. Every detection's
magnitude is calculated, using the following formula:
\begin{equation}
R = 27.36 - 2.5 \log{f_5 / t} - 0.09 X ,
\end{equation}
where $f_5$ corresponds to the flux in a 5-pixel aperture, $t$ is the
time in seconds and $X$ is the airmass. This equation accounts for an
average 0.34~mag aperture correction between the known magnitude of a
synthetic object and its magnitude measured with a 5-pixel
aperture. The search algorithm requires an object to be found in all
three exposures giving three independent magnitude measurements that
we average to obtain the results shown in Table \ref{tab:fits}. The
errors given on the magnitude values correspond to the error on the
flux.

\begin{figure}[ht]
\epsscale{1.0} \plotone{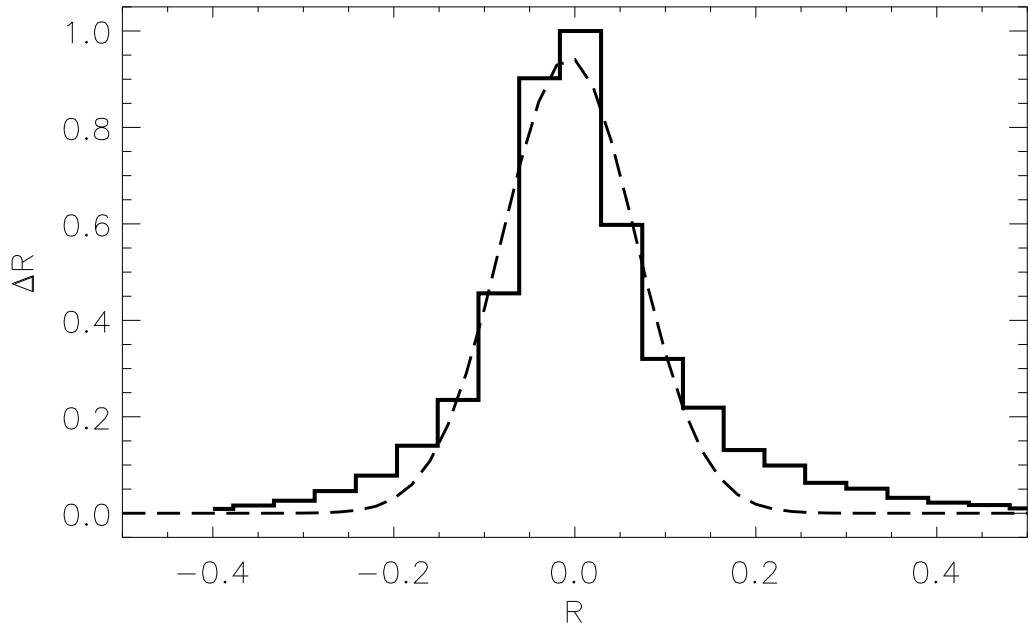}
\caption{\label{fig:magerr} Histogram of the magnitude error
  ($\Delta$R) as a function of R magnitude for all implanted
  objects. The error is defined as the difference between the
  implanted and measured magnitudes for the synthetic population. The
  dashed line is a gaussian of width $\sim$0.1 mag.  }
\end{figure}

\begin{figure}[ht]
\epsscale{1.0}
\plotone{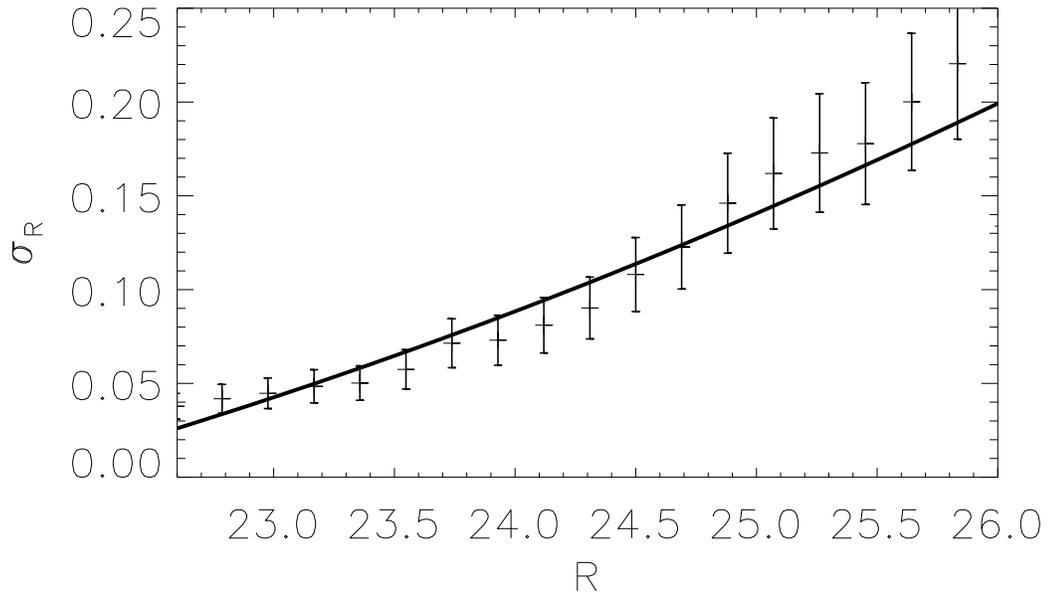}
\caption{\label{fig:magerrofmag} The error in magnitude for synthetic
  objects as a function of magnitude is shown for different magnitude
  bins. The error is defined as the FWHM of the best-fit gaussian to
  the histogram of errors for all objects in each bin. The error bars
  correspond to the calculated uncertainty of the FWHM. The curve is a
  quadratic fit to the data and defines the error estimate used for
  $\Delta$R$_{mag}$ in Table \ref{tab:fits}.  }
\end{figure}

In Figure \ref{fig:magerr} we plot the photometric errors, showing
them to be $\sim$0.1 mag. The magnitude dependence of the uncertainty
is shown in figure~\ref{fig:magerrofmag}. We estimate the
uncertainties empirically, calculating the standard deviation as a
function of magnitude. We then fit a second degree polynomial,
overlayed in Figure~\ref{fig:magerrofmag}. This estimate is shown for
each real object as $\Delta_{Rmag}$ in Table \ref{tab:fits}.

\begin{figure}[ht]
\epsscale{1.0} \plotone{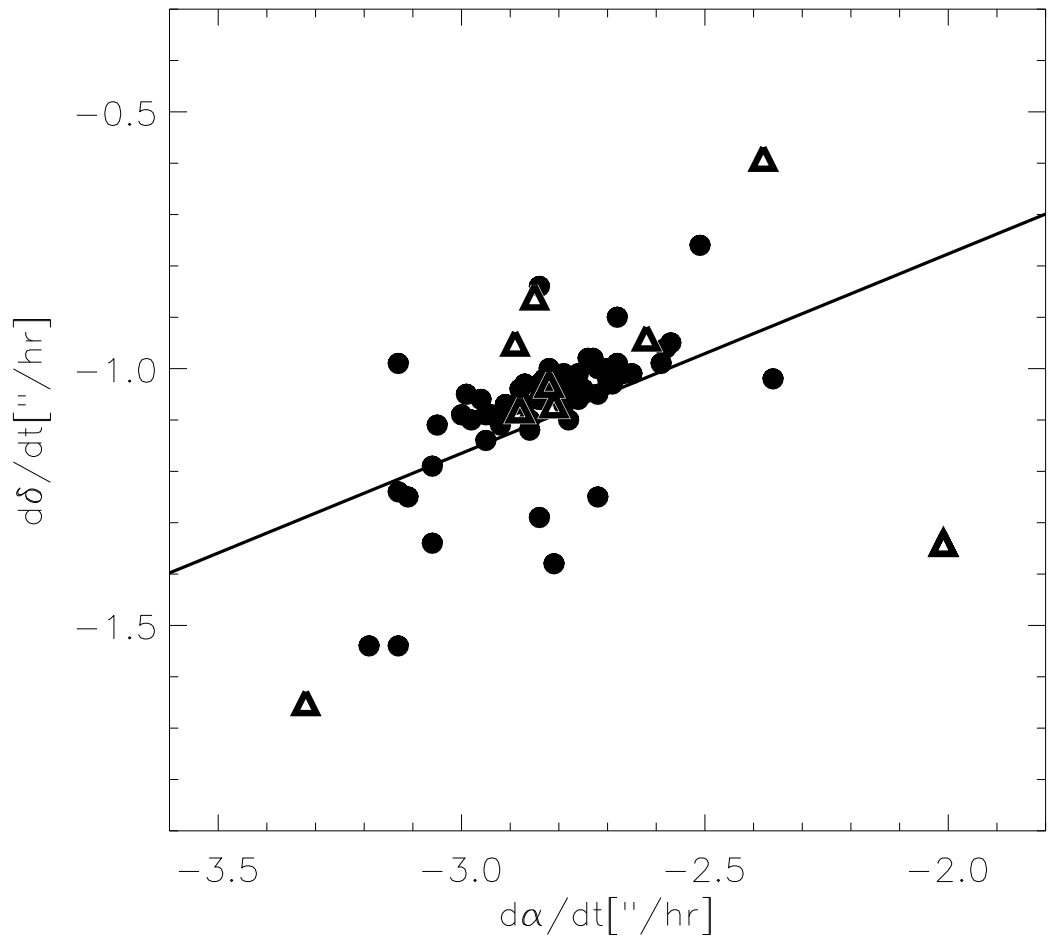}
\caption{\label{fig:xyerr} Rate of motion in the sky for every
  TNO. Objects observed only on one night only are represented by
  triangles. The ecliptic motion is overplotted as a solid line.  }
\end{figure}

We can accurately approximate the apparent motion of a TNO over 24
hours as a straight line with a constant rate. We include the measured
right ascension and declination rates in Table \ref{tab:fits}. The
apparent motion of our objects compared with that which parallels the
ecliptic is plotted in \fig{fig:xyerr}.

Near opposition, the change in the rate or direction of motion over 24
hours is negligible, making it easy to predict where the real objects
would be on the second night. However, nine objects were not found on
the second night. Our method is only $\sim90\%$ efficient for the
brightest objects on the first night, with 10\% lost to blending with
field stars. It is expected that more than 10\% of the objects will be
lost on the second night, because of confusion with stationary sources
and because they are more likely to move outside the field of view
over 24 hours.

We used the observations on the second night to improve the distance
determination when possible. We use the \citet{Bernstein.2000}
{\it Orbfit} routines to estimate plausible orbital elements assuming
there's no acceleration in the direction tangential to the plane of
the sky. For a 24-hout arc, this results in a $\sim$7\% accuracy in
the barycentric distance ($d_{bari}$). For a single night observation
of objects the error on the distance could be unbound. However, since
the observations were taken near opposition we are able to readily
estimate heliocentric distances ($d_{par}$) from the ``parallactic
motion''. We assume that the observations are taken exactly at
opposition and that the orbits are circular. This distance estimate is
not as reliable as $d_{bari}$ but it serves as a consistency check.

\subsection{Statistical Analysis}\label{sec:stat}
The probability of our data ($D$) given a model for the intrinsic
population ($M$) is denoted $P(D|M)$=$L(M)$, where $L$ is the
likelihood function. We consider the data in our survey as a
collection of $N$ detections with measured magnitudes. As derived in
\citet{Schechter.1976} if $g(m)dm$ is the expected number of
detections between $m$ and $m+dm$, then the likelihood of a set $m_i$
where $i$=$1,\cdots,N$ is:
\begin{equation}
 L(M) = exp [-\int\limits_{-\infty}^{\infty}
g(m)dm ] \prod\limits_{i=1}^N g(m_i)dm.
\end{equation}
 We are interested in characterizing $g(m)$. As described in
\citet{Bernstein.2004}, we can think of $g(m)$ as being the
probability of detecting an object and assigning it a magnitude $m$
given the survey characteristics and the real distribution of objects
on the sky. We consider an intrinsic differential surface density of
objects $\sigma$ that only depends on magnitude and is constant over
the observed area as the model $M$. For a survey with an efficiency
function $\eta$, a function of magnitude only, we can write $\int
\limits _{-\infty}^\infty g(m) dm = \Omega \int \eta(m) \sigma(m) dm
$, where $\Omega$ is the solid angle of the survey.

The likelihood of a model for the differential surface density
$\sigma(m)$ is then given by:
\begin{eqnarray}\label{eqn:lik}
  L(\sigma) = e^{-\Omega \int \eta(m) \sigma(m) dm}
  \prod \limits_i \int l_i(m) \eta(m) \sigma(m) dm.~~~
\end{eqnarray}
This is the probability of finding each object in the set of
observations at its measured magnitude, scaled by the probability of
not finding anything else. The function $l_i(m)$ is the probability an
object is given a magnitude $m_i$ given its intrinsic magnitude is
$m$.

If we consider the efficiency function and model it as relatively
linear over the magnitude uncertainty of an observation we can
approximate our likelihood function as follows:
\begin{eqnarray}\label{eqn:likeapprox}
  L(\sigma) = e^{-\Omega \int \eta(x) \sigma(x) dx}
  \prod \limits_i \eta(m_i) \sigma(m_i)
\end{eqnarray}
This is extremely useful when dealing with a large number of objects
and surveys. We compared the behavior of both exact and approximate
likelihood functions with our data and found no noticeable
differences.

If we want to sample the likelihood function over its parameter space
or calculate the total likelihood of a model we need to consider
priors. That is, the probability of a parameter $q$ given a certain
model $M$, $P(q|M)$. These priors reflect any knowledge we have over
the value of a parameter previous to our survey. We chose priors that
reflect the least previous knowledge into the analysis. We chose
uniform functions between limits set by our survey, indicating our
ignorance of those parameters. The total probability of a model is:
\begin{eqnarray}\label{eqn:likeofmod}
  \mathcal{L}(\sigma) =
  \int P(q|\sigma) L(\sigma,q) dq
\end{eqnarray}
We can compare two competing models using their total likelihoods by
computing the odds ratio:
\begin{eqnarray}\label{eqn:occam}
  O_{21}& = &\frac{P(\sigma_2|D)}{P(\sigma_1|D)} =
  \frac{P(\sigma_2)}{P(\sigma_1)} \frac{P(D|\sigma_2)}{P(D|\sigma_1)}
  = \frac{\mathcal{L}(\sigma_2)}{\mathcal{L}(\sigma_1)} ~~~
\end{eqnarray}
The last equality holds if we do not have a good reason to prefer ``a
priori'' any of the two models. The ratio of the total likelihoods is
called Bayes factor.

\subsection{Single Power Law Model}
One of our goals is to determine whether the results of our survey
indicate that the cumulative surface density can be modeled by a
single power law distribution (SPL) or if the data favor a more
complicated model. We use a likelihood analysis to investigate
this. 

The likelihood function is related to both the detection
efficiency of the survey and the differential surface density
$\sigma(R)$. The important observation for the analysis is the number
of objects we detect brighter than a given magnitude, namely the
cumulative surface density: $\Sigma(R)$=$\int_{-\infty}^R \sigma(x)
dx$. We use the likelihood function given by eq.~\ref{eqn:lik}, with
$\Omega$=$2.83\sqdeg$ and $\eta(R)$ given by eq.~\ref{eqn:eff}. 

For every object we model its photometric uncertainty using the
analytical model we considered previously, a gaussian ($l_i$) around
its measured magnitude (see figure \ref{fig:magerrofmag}).

\begin{figure}[ht]
\epsscale{1.0} \plotone{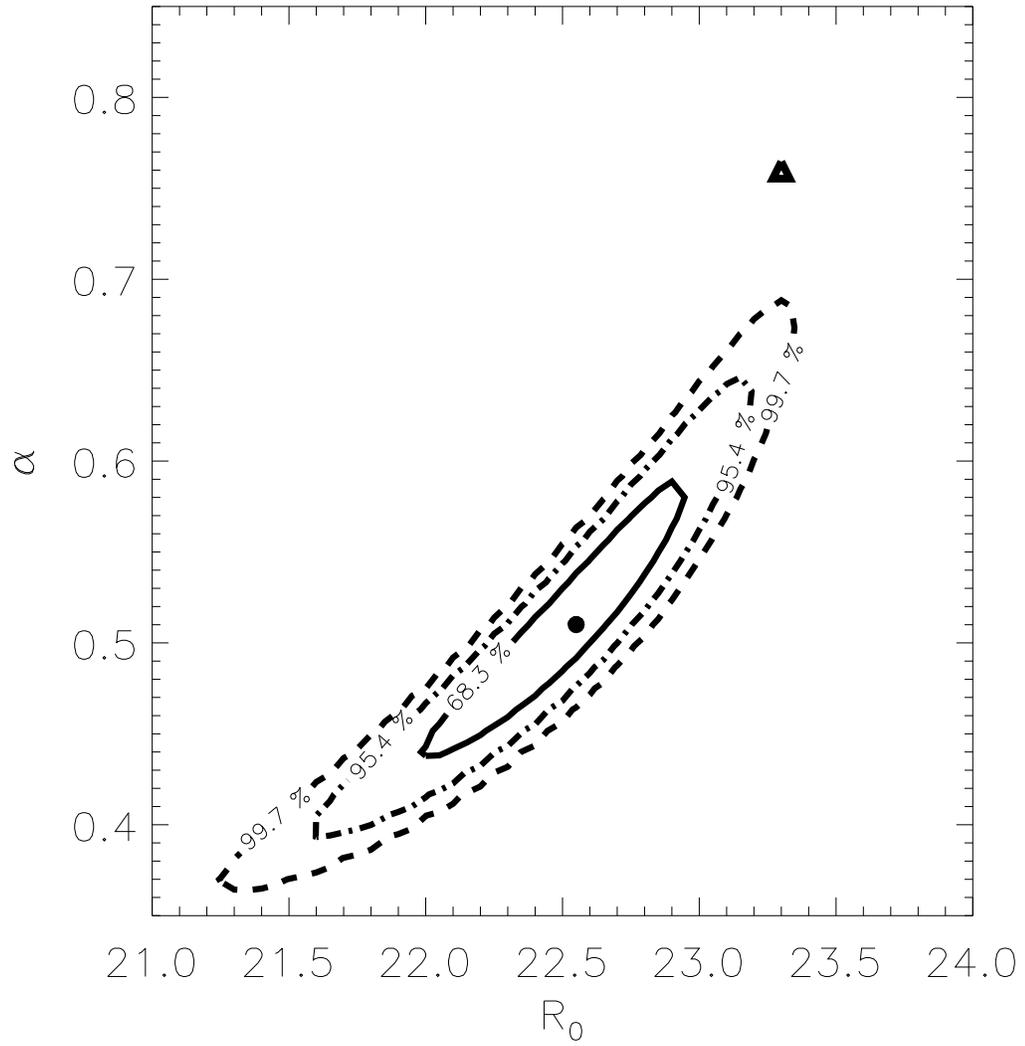}
\caption{\label{fig:like1} Contours of the SPL likelihood
  function. The maximum likelihood point is marked with a dot
  ($\alpha$=0.51, $R_0=$22.6). Marked with a triangle is the best
  value for the parameters based on \citet{Petit.2006},
  ($\alpha$=0.76, $R_0$=23.3). This shows the discrepancy between our
  result and that of \citet{Petit.2006}.  }
\end{figure}

The single power law model is written as follows:
\begin{eqnarray}\label{eqn:sig1}
  \sigma_1(R,\alpha,R_0) = \alpha \ln(10) 10^{\alpha (R-R_0)}
\end{eqnarray}
In Figure \ref{fig:like1} we plot the SPL likelihood as a function of
$R_0$ and $\alpha$. The previously accepted values for the SPL
parameters ($\alpha$=0.76, $R_0$=23.3) \citep{Petit.2006} are in
strong disagreement with our data, lying well outside our 3-$\sigma$
confidence region. Most of the surveys that have consistently measured
a slope of $\alpha$$\sim$0.7 for the cumulative distribution have
brighter limiting magnitudes \citep{Gladman.1998, Petit.2004,
Petit.2006}. The exceptions are \citet{Gladman.2001} and
\citet{Fraser.2008}, who quote a magnitude limit of $R$=25.9 and
$R$=25.6 respectively. \citet{Bernstein.2004} performed a search
complete to $R$=28.5. They discovered far too few objects to be
consistent with a SPL.

\begin{figure}[ht]
\epsscale{1.0} \plotone{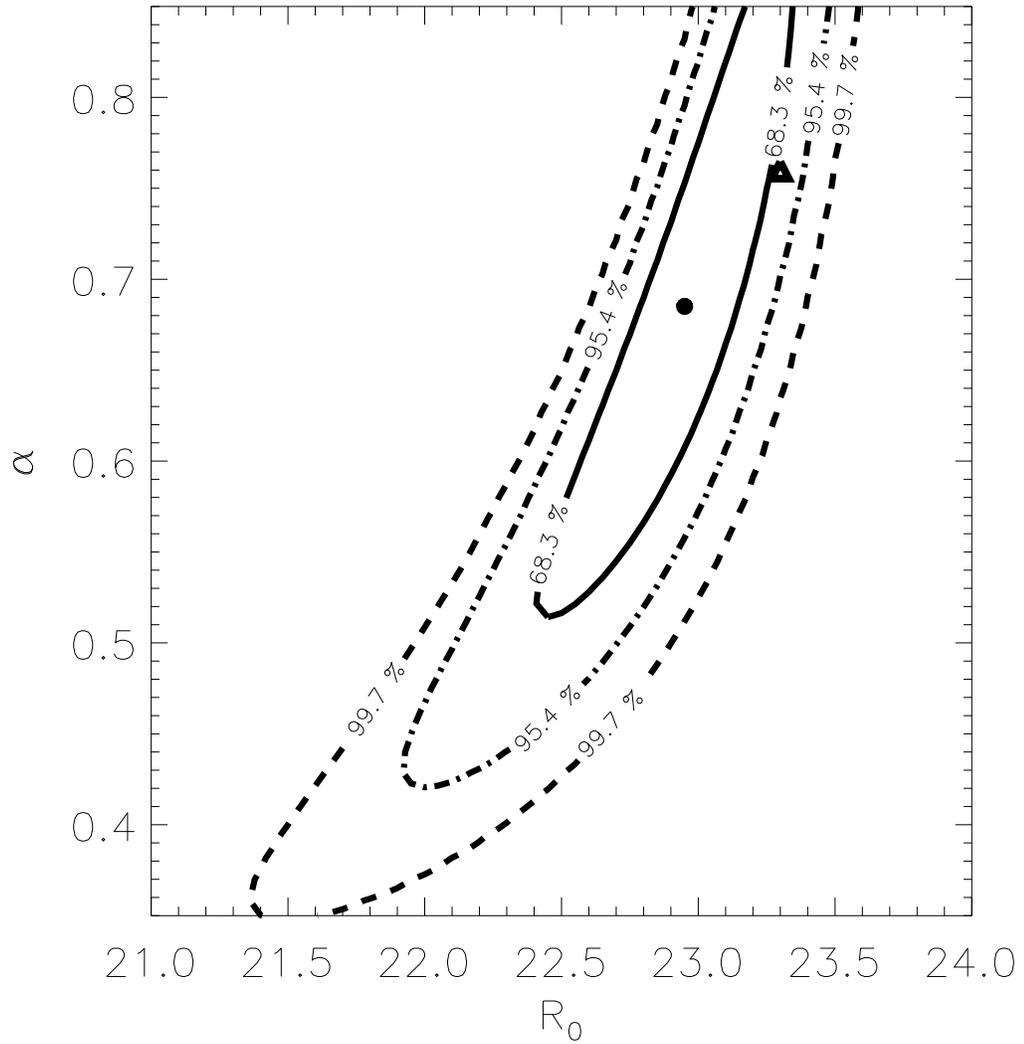}
\caption{\label{fig:like1_cut} Contours of the SPL likelihood function
  for our sample limited to $R\leqslant$24.5. The maximum likelihood
  point is marked with a dot ($\alpha$=0.69, $R_0$=23.0). Marked with
  a triangle is the best value for the parameters based on
  \citet{Petit.2006}. Both results consistent with each other. This
  shows that our survey agrees with previous surveys if we consider
  the only the range of magnitudes to which those surveys are sensitive.  }
\end{figure}

We check that our bright end sample is consistent with the
previous surveys with shallower limiting magnitudes. In
Figure~\ref{fig:like1_cut} we plot our sample's likelihood function
after 
imposing an artificial efficiency limit at $R$=24.5. The power law
index is clearly consistent with the \citet{Petit.2006} result and it
shows our sample does not deviate from the SPL behavior observed by
others for magnitudes brighter than $R$$\sim$24.5.

\begin{figure}[ht]
\epsscale{1.0}
\plotone{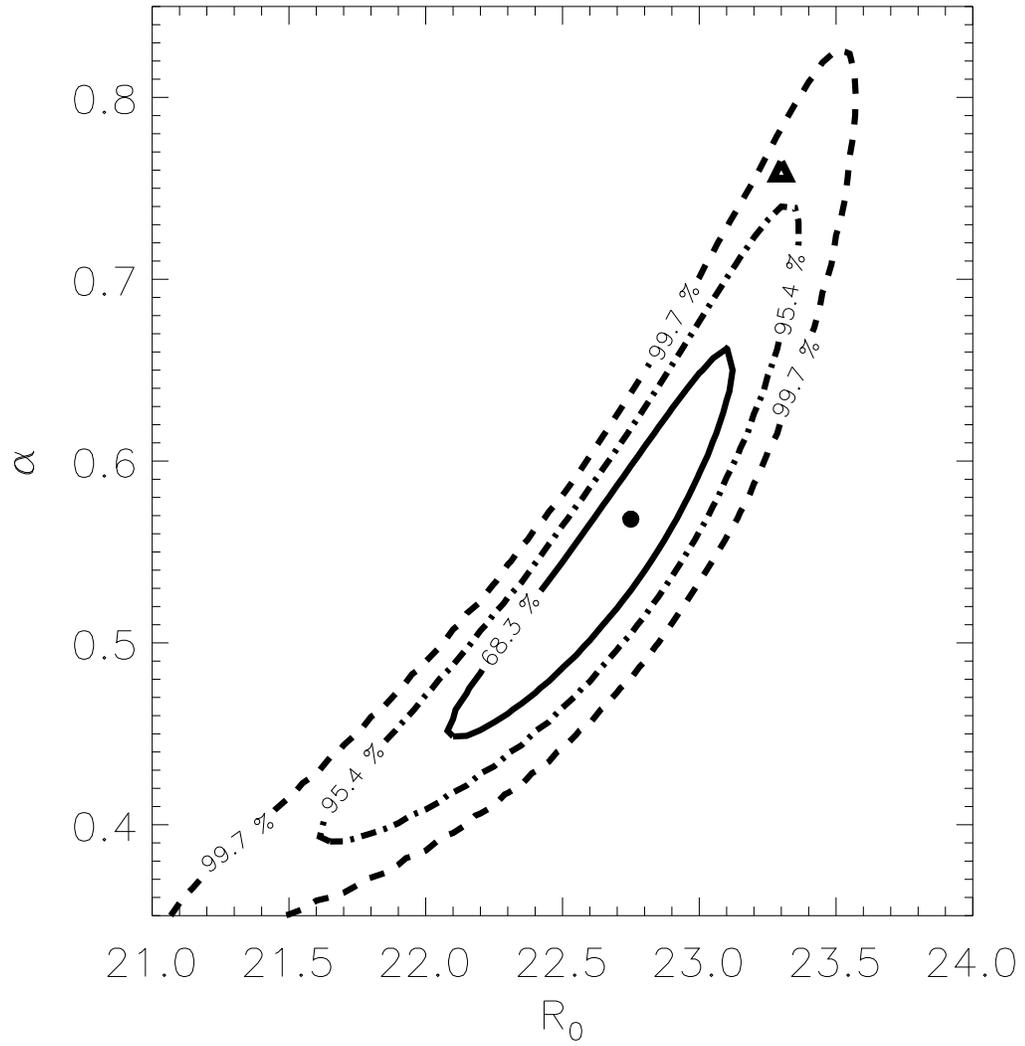}
\caption{\label{fig:like1_cut70} Contours of the SPL likelihood
  function for our sample limited to $R\leqslant$25.2, where our
  survey is $70\%$ efficient. The maximum likelihood point is marked
  with a dot ($\alpha$=0.57, $R_0$=22.8) and the triangle is the
  \citet{Petit.2006} result. We see that both results are inconsistent
  at more than a 2-$\sigma$ level. This demonstrates that our result does not
  rely on the detection of objects at magnitudes where our detection
  efficiency is declining.  }
\end{figure}

To show that the deviation from a SPL at fainter magnitudes is not an
artifact of our efficiency function, we repeated the experiment but
instead imposed an artificial break at $R$=25.2, where our survey is
$70\%$ as efficient as its maximum efficiency. The result can be seen
in $\Fig{fig:like1_cut70}$, it shows the \citet{Petit.2006} result is
rejected at the $2$-$\sigma$ level.

\subsection{Double Power Law Model}
Now that we have shown that our results are not well modeled by an
SPL, we test a more complicated model. Any model that includes a break
in the surface density distribution will have more free parameters
than an SPL. Alternatives with three and four parameters were tried by
\citet{Bernstein.2004} to explain the aforementioned under-abundance
of detections. We will focus on the ``double power law'' (DPL) model,
the harmonic mean of two different power laws. Though a model with
three parameters would be easier to implement, it does not provide the
immediate insight into the TNO population that the DPL provides. The DPL
has four free parameters, allowing two different asymptotic power law
behaviors for the distribution (that can be linked to the size
distribution of small and large objects), a break in the luminosity
distribution, and a differential density constant.

\begin{figure}[ht]
\epsscale{1.0} \plotone{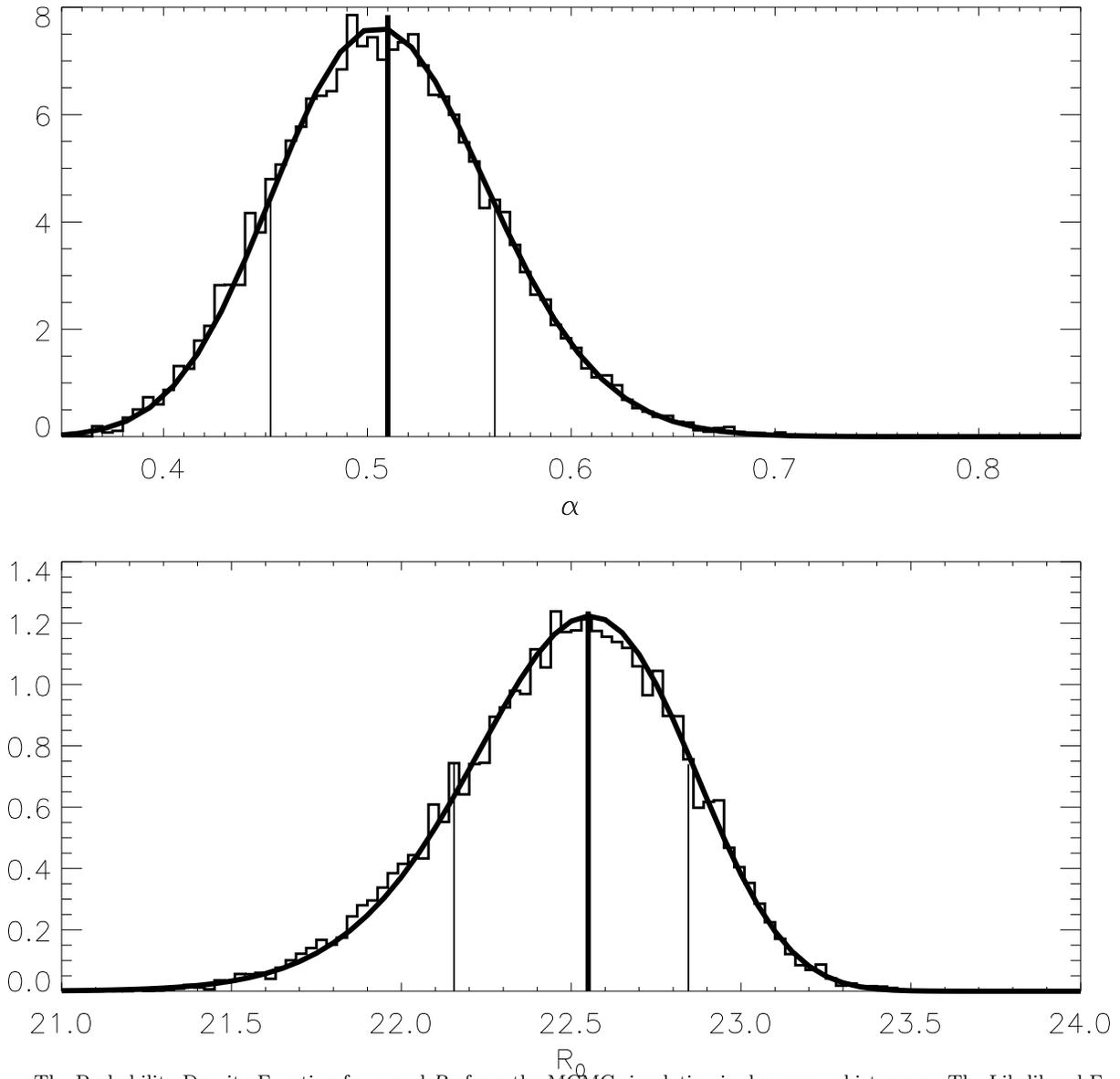}
\caption{\label{fig:compmcmc} The Probability Density Function for
  $\alpha$ and $R_0$ from the MCMC simulation is shown as a
  histogram. The Likelihood Function in \fig{fig:like1}, shown as the
  marginal probability over each parameter is plot as the solid curve.
  The solid, heavy line indicates the global maximum obtained by the
  MCMC run and the thin lines indicate the 1-$\sigma$ credible region
  of the parameter, inside of which we find 68.3\% of the probability.
  }
\end{figure}

The larger number of parameters makes the likelihood function more
difficult to sample, thus we use a Markov Chain Monte Carlo (MCMC)
approach (for an MCMC review see \citealt{Tegmark.2004}). We use a
Metropolis-Hastings algorithm to sample the likelihood function with a
gaussian proposal distribution. The parameters were set to yield a
$\sim$25\% acceptance rate. We considered a run of 100,000
iterations. To check for consistency we tried different initial
conditions and compared the results, no disagreement was found. We
also checked the performance of our MCMC code with the SPL model. In
\Fig{fig:compmcmc} we show the marginalized probability for both
parameters $\alpha$ and $R_0$ from MCMC and the exact result. There is
evident agreement between the two approaches.

The DPL likelihood function is obtained by replacing the corresponding
surface number density (\Eq{eqn:sig2}) in the likelihood function
(\Eq{eqn:lik}) with:
\begin{eqnarray}\label{eqn:sig2}
  \sigma_2(R) & = & C
   \left[ 10^{-\alpha_1 (R-23)}+10^{(\alpha_2-
   \alpha_1)(R_{eq}-23)-\alpha_2 (R-23)} \right] ^{-1} \nonumber,\\
  C & = & \sigma_{23} (1+10^{(\alpha_2- \alpha_1)(R_{eq}-23)})
\end{eqnarray}

\begin{figure}[ht]
\epsscale{1.0} \plotone{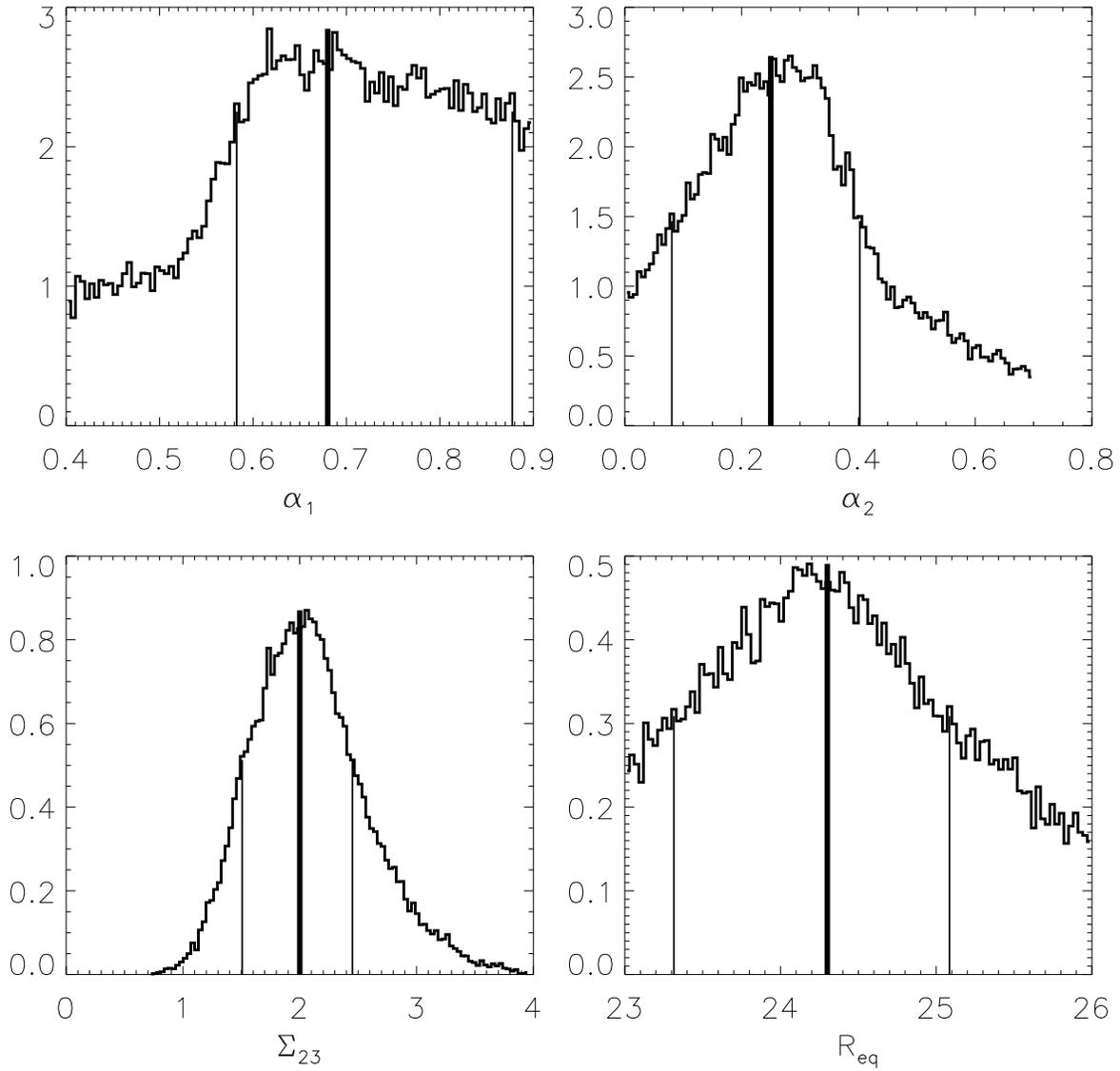}
\caption{\label{fig:like2} The DPL likelihood for our survey as a
  function of all parameters is shown in each window. The most likely
  parameters and their 1-$\sigma$ confidence regions, represented by
  the solid, heavy line and the two thin lines, are
  $\alpha_1$=$0.7_{-0.1}^{+0.2}$, $\alpha _2$=$0.3_{-0.2}^{+0.2}$,
  $\sigma_{23}$=$2.0_{-0.5}^{+0.5}$ and $R_{eq}$=$24.3_{-0.1}^{+0.8}$.
  }
\end{figure}

In \fig{fig:like2} we show the DPL likelihood as a function of the
bright-end slope $\alpha_1$, the faint-end slope $\alpha_2$, the value
of the surface number density at $R=23$ $\sigma_{23}$ and the break
magnitude $R_{eq}$. All parameters but $\alpha_1$ are well constrained
by the data. Given the small number of bright TNOs detected in our
survey, the limited constraint on $\alpha_1$ is not surprising.

\subsection{Cumulative Number Density}
\begin{figure}[ht]
\epsscale{1.0} \plotone{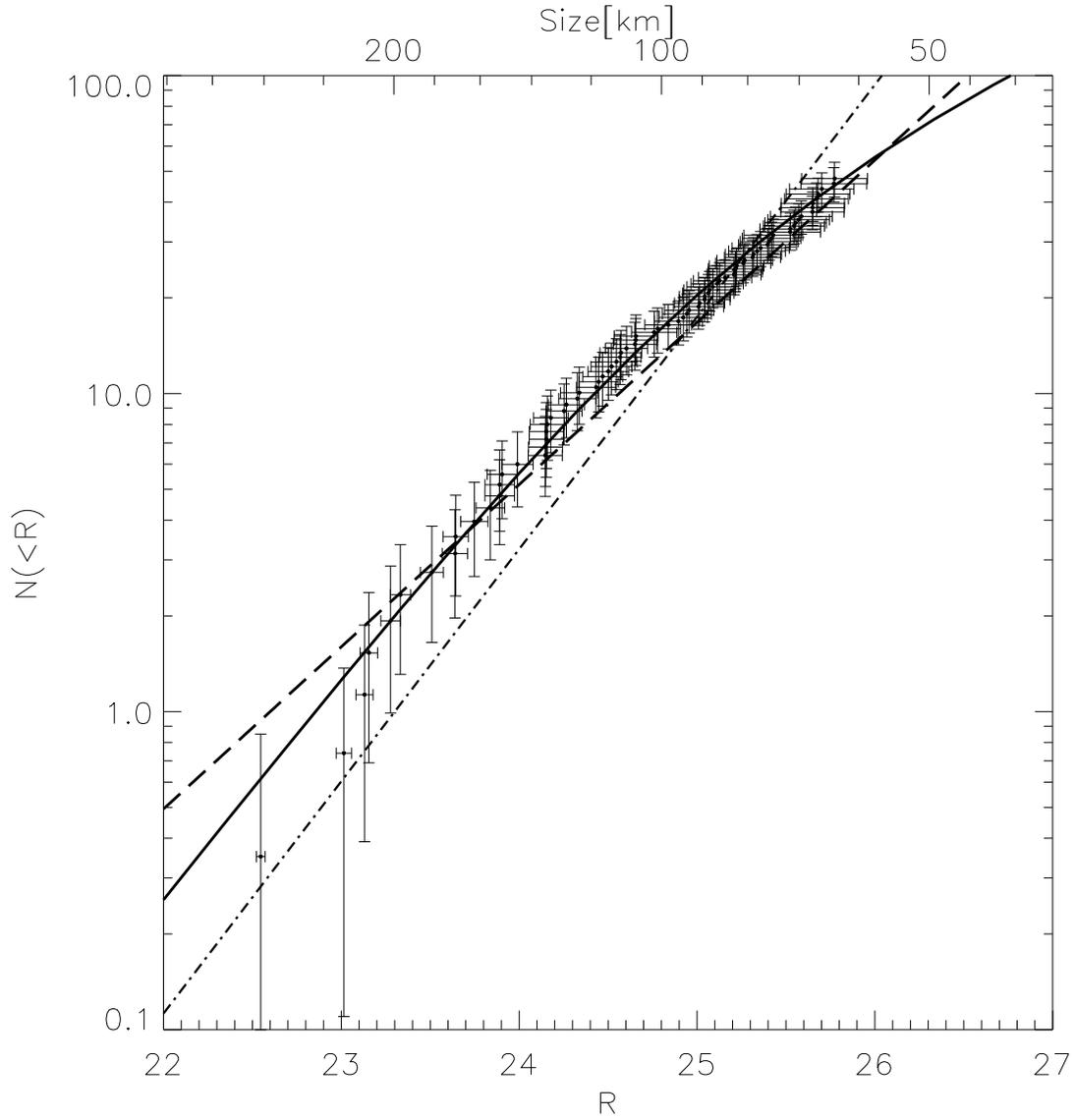}
\caption{\label{fig:cumfunc} The cumulative number density for our
  survey. The best previous model is plotted in the short-dashed
  line. Our most likely solution for the single power law is plotted
  in the long-dashed line. The best DPL fit is shown as a solid
  line. The quoted size corresponds to an object at $42\au$ and 4\%
  albedo.  }
\end{figure}

Using the detection efficiency (\Eq{eqn:eff}) we can estimate the
number of objects we missed for each object found. We construct a
cumulative function of the unbiased population plotting each object
individually, representing with its detection a number of objects with
similar magnitudes. Since we are plotting a cumulative function, the
errors are correlated (See \Fig{fig:cumfunc}).

We go on to compare the total likelihood of both models, as described
in \S~\ref{sec:stat}. A simple way of doing this is to examine the
goodness-of-fit of the cumulative number density. Figure
\ref{fig:cumfunc} shows the data and the best solution for the single
and double power law cases. Note that those power laws correspond to
the cumulative number densities, $\Sigma_1(R)$=$10^{\alpha (R-R_0)}$
and $\Sigma_2(R)$=$\int_{-\infty}^R \sigma_2(x) dx$.

It is expected that a DPL gives a better fit to the data than a SPL
model. The question is whether this better fit overcomes the increased
complexity in the model. This can be answered calculating the quotient
of the total bayesian probabilities of the models (Bayes factor,
details in Appendix \ref{sec:stat}). If the total probability for a
given model is larger than another then it is preferred. Using the
results of the MCMC simulations we compute this factor. The resulting
total probabilities depend on suitable priors, that reflect our
ignorance on the parameters. We selected uniform priors for all our
variables. For the SPL we chose $\alpha~\epsilon~[0.35, 0.85]$,
$R_0~\epsilon~[21.0, 24.0]$, while the DPL priors were uniform,
$\alpha_1~\epsilon~[0.5, 1.0]$, $\alpha_2~\epsilon~[0.1, 0.7]$,
$\sigma_{23}~\epsilon~[0.5, 5.0]$, $R_{eq}~\epsilon~[23.0, 26.0]$. The
calculated Occam's factor is $O_{sd}$=26, meaning that a DPL model is
more likely to be a better representation for the brightness
distribution of our data.

\subsection{Other Surveys}
\citet{Bernstein.2004} combined the results of their {\it HST} survey
with those of \citet{Chiang.1999, Gladman.1998, Allen.2002,
Trujillo.2001, Larsen.2001, Trujillo.2003}. We include most of the
objects listed in that work and those conducted
since. $\tab{tab:surv}$ differs from \citet[Table 2]{Bernstein.2004}
in the exclusion of the two widest searches and the inclusion of two
newer surveys \citep{Petit.2006,Fraser.2008}, as well as ours. We excluded the
two surveys because of the complexity in establishing the searched
area near the ecliptic. For the sake of comparison with
\citet{Bernstein.2004} we use the same criteria regarding detected
objects as well as the caveats provided therein. We included surveys
for which the location of the searched area, effective area of the
search, magnitude at which the efficiency drops by 50\% must be
given.  We include objects that have an observed magnitude where their
efficiency function is more than $15\%$ the maximum efficiency of the
survey. We point out that all our detections satisfy this requirement.

\begin{figure}[ht]
\epsscale{1.0} \plotone{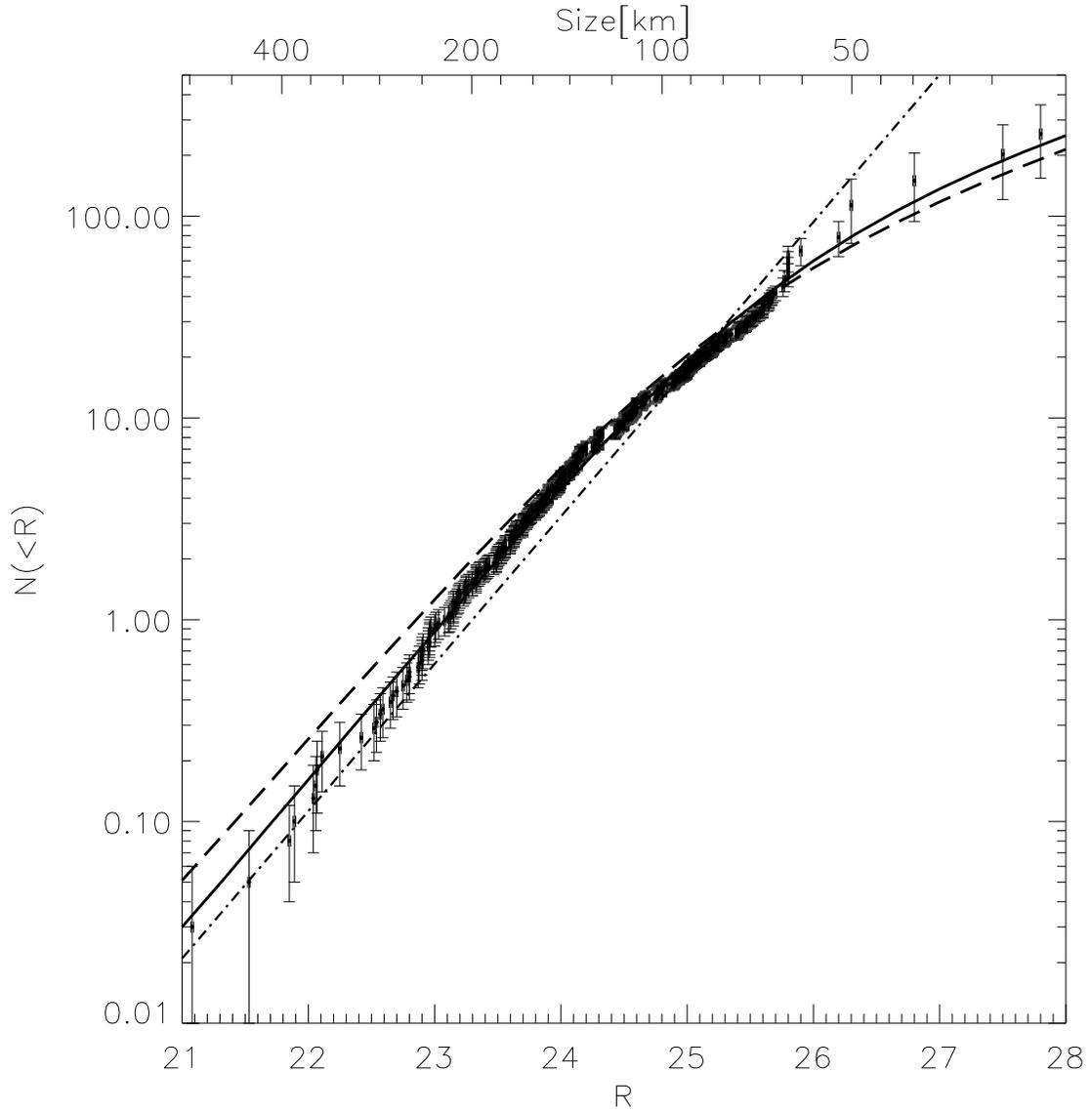}
\caption{\label{fig:cumfuncall} The cumulative number density for all
  surveys in Table \ref{tab:surv}. The best previous model is plotted
  in the black dashed line. Our most likely double power law is
  plotted in the long-dashed line. The most likely DPL (see
  $\Fig{fig:like2all}$) considering all surveys is plotted as a full
  line. The apparent bump in density at around $R$$\sim$25.8
  corresponds to 5 objects in \citet{Gladman.2001}. }
\end{figure}

\begin{figure}[ht]
\epsscale{1.0} \plotone{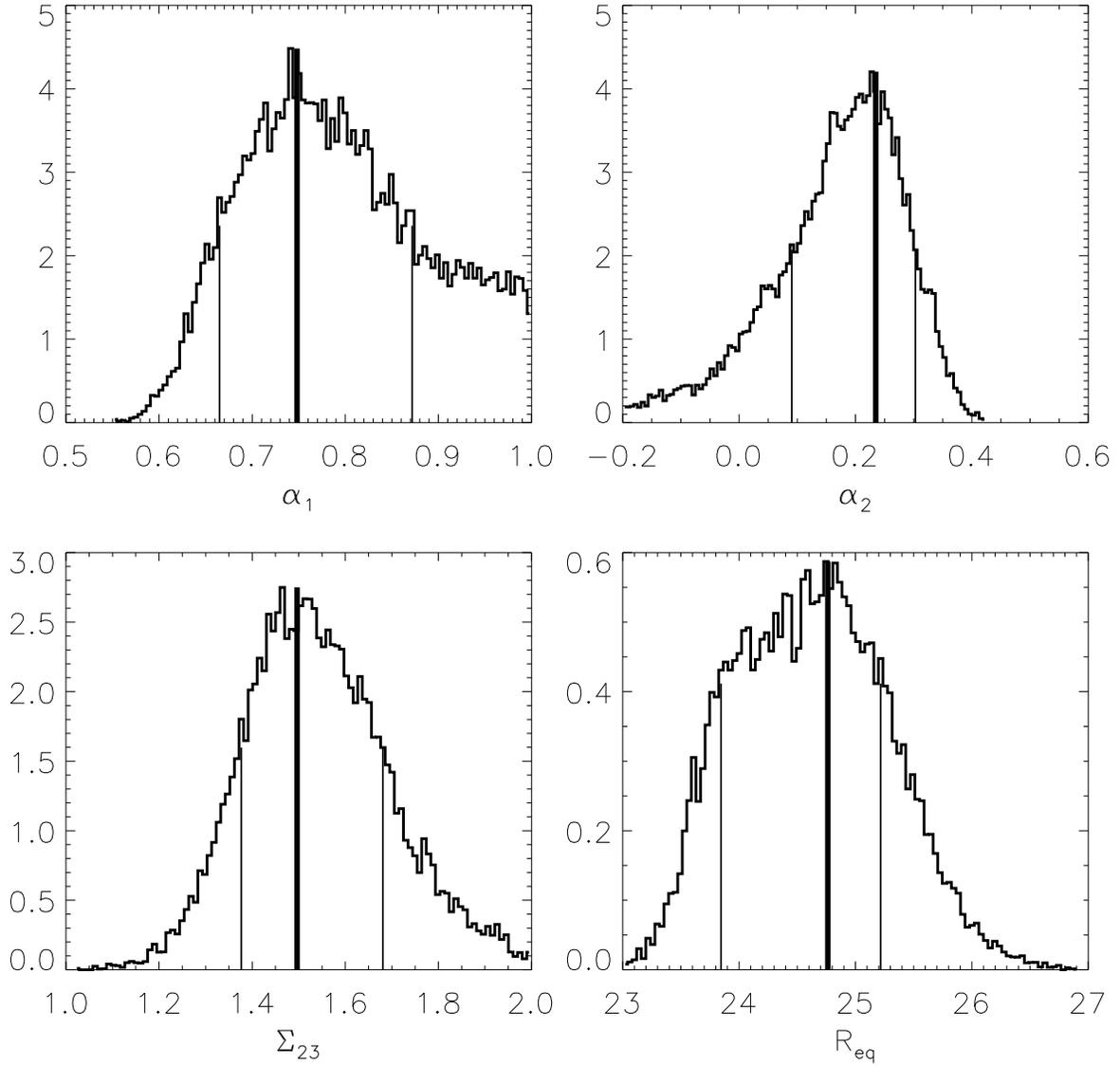}
\caption{\label{fig:like2all} The DPL likelihood function marginalized
  over each parameter for all the surveys in Table \ref{tab:surv}. We
  see the maximum and 68\% confidence region. The most likely value
  for each parameter and 1-$\sigma$ confidence limits are:
  $\alpha_1$=$0.75_{-0.08}^{+0.12}$,
  $\alpha_2$=$0.23_{-0.14}^{+0.07}$,
  $\sigma_{23}$=$1.50_{-0.12}^{+0.18}$ and
  $R_{eq}$=$24.8_{-0.9}^{+0.5}$.  }
\end{figure}

We are interested in computing the likelihood of a model given the
data from each survey. For this we only need the list of objects that
meet our criteria, an estimate of the efficiency function, the
surveyed area, and a way to translate all measurements to the red
filter R for each survey. We use the approximation given in
(\Eq{eqn:likeapprox}). Figure \ref{fig:cumfuncall} shows the 333
objects that we considered. It shows the existence of a very
pronounced lack of detections at faint magnitudes. Our likelihood
analysis is summarized in $\Fig{fig:like2all}$ with 1-$\sigma
$confidence limits for the parameters.

An interesting aspect of our search is that the data has been
available since August 2003.  Our survey's most likely distribution
expects $\sim$12 detections for the {\it HST} field while $3$ were
found. This provides independent support to the existence of a break
in the TNO luminosity function.

\subsection{Classical \& Excited Population}
We use the criteria in \citet{Bernstein.2004} to identify
``Classical'' and ``Excited'' objects. TNOs with distance at discovery
$d$ between $38\au$ and $55\au$ and inclination $i\leq5\deg$ are
considered ``Classical'' and the rest are considered ``Excited''. In
Table \ref{tab:surv} we list each survey with the corresponding number
of TNOs in each category.

This survey was considered by itself and together with the surveys in
Table \ref{tab:surv}. We investigated how does the DPL luminosity
function change when applied to the different populations. We repeated
the MCMC analysis for both populations and for our survey and the
combined survey. We also considered the priors used in
\citet{Bernstein.2004}, $-0.5<\alpha_1,\alpha_2<1.5$ to constrain the
parameter space.

The results of the MCMC simulations are summarized in Table
\ref{tab:params}. These results are very similar to those by
\citet{Bernstein.2004}. However, we have included three new surveys
(This survey and those by \citealt{Petit.2006, Fraser.2008}), two of
which (This survey, \citealt{Fraser.2008}) sample magnitudes fainter
than $R$$\sim$25.5 where excited objects were specially under-sampled.

\subsection{Size, Distance \& Inclination Distribution}
The size distribution is closely related to the distribution of
apparent magnitudes. It is customary to assume all objects are located
at the same distance and that the size distribution is a single power
law and hence the cumulative brightness distribution is also a power
law. The parameters of the two distributions are related by
$q=5\alpha+1$, where $q$ is the exponent of the differential size
distribution ($dn=D^{-q}dD$) and $\alpha$ is the exponent of the SPL
cumulative luminosity function.

\begin{figure}[ht]
\epsscale{1.0} \plotone{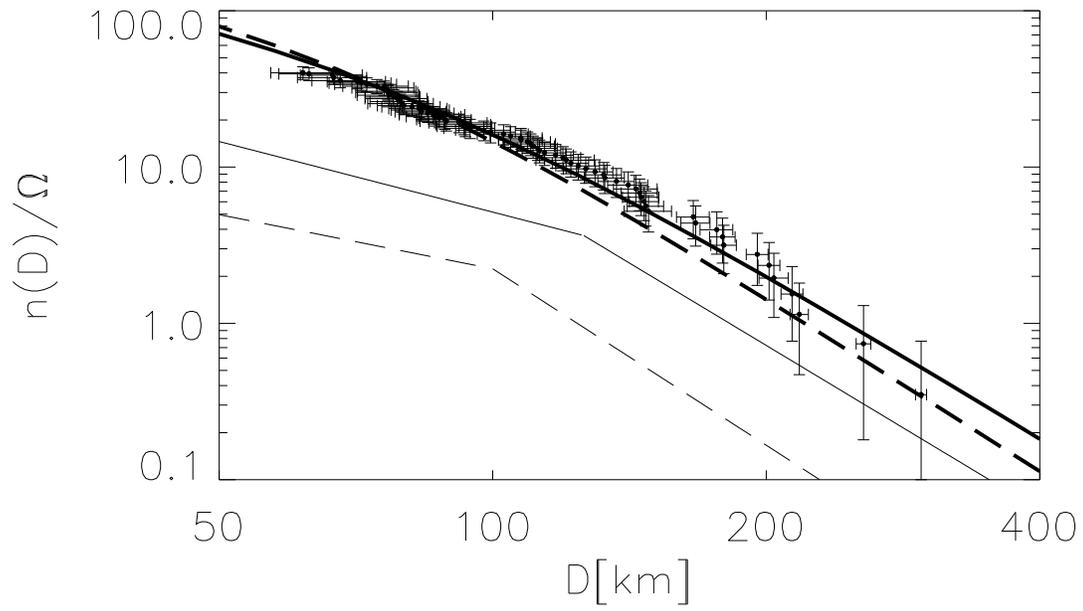}
\caption{\label{fig:sizedist} Number of TNOs observed in 1$\sqdeg$ as
  a function of size. The solid line shows the model based on our
  survey. The dashed one is the model that considers all surveys. Both
  models are properly scaled to match the density observed in our
  survey.  }
\end{figure}

With our rough distance estimates and assuming a 4\% albedo for TNOs,
we can compute the real size distribution of the objects in our survey
(we adopt $m_R=-27.6$ for the R band magnitude of the Sun). In Figure
\ref{fig:sizedist} we show the cumulative size distribution for our
survey. However, the typical error in distance $\sim$7\% translates
into a 0.3 magnitude photometric error, triple the median photometric
error in our survey (see Figure \ref{fig:magerr}). Thus, instead of
repeating the statistical analysis for the size distribution directly,
we transform our luminosity function into a size distribution assuming
all objects are located at $42\au$. The best DPL fits for the
luminosity function are plotted as a function of size. The solid line
is the fit to this survey and the dashed line corresponds to the fit
to the surveys in Table \ref{tab:surv}. We also consider a toy model
based on the DPL; it corresponds to two power laws with index
$q_1=5\alpha_1+1$ and $q_2=5\alpha_2+1$ that are joined at the size
for which an object at $42\au$ would be observed to have magnitude
$R_{eq}$. We plot the cumulative function of the toy models for both
DPLs to show the asymptotic behaviors as a light solid line and a
light dashed line respectively, both are arbitrarily offset vertically
for clarity. In Figure \ref{fig:sizedist} there is a clear agreement
between the real size distribution and the fit for the DPL models
indicates that the assumption that all TNOs are at the same distance
is justified.

\begin{figure}[ht]
\epsscale{1.0} \plotone{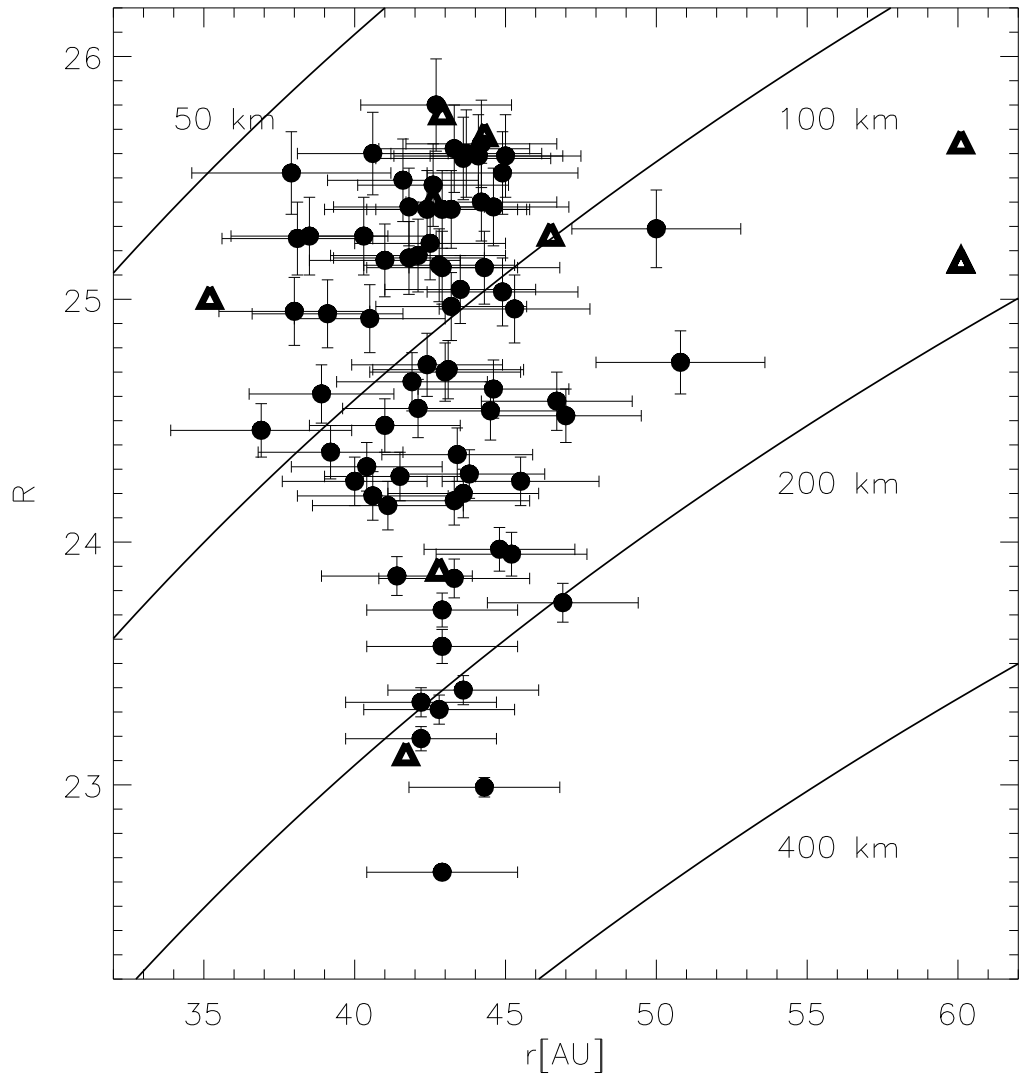}
\caption{\label{fig:dist} Magnitude and distance for all 82 TNO's
  found. The black dots are objects observed in both nights and the
  triangles are those with only one night's observation.  We assume a
  4\% albedo to plot the constant size curves for 100, 200 and 400 km
  in black.  }
\end{figure}

In Figure \ref{fig:dist} we plot the distance and magnitude for each
object. The distance corresponds to $d_{bari}$ in Table \ref{tab:fits}
with the exception of those objects that were not recovered on the
second night for which we plot the circular orbit approximation
($d_{par}$). We consider only the subset of 73 objects with 24 hour
arcs data, with a distance error of $~5\%$.

All but two objects are located at less than $50\au$ from the Sun,
although we are able to detect $D=250\km$ TNOs at distances of
$80\au$, with 50\% efficiency. This lack of distant detections has
been noted previously \citep{Allen.2001, Trujillo.2001,
Bernstein.2004} with the recurrent hint that there is an ``edge'' to
the Kuiper Belt.

\begin{figure}[ht]
\epsscale{1.0} \plotone{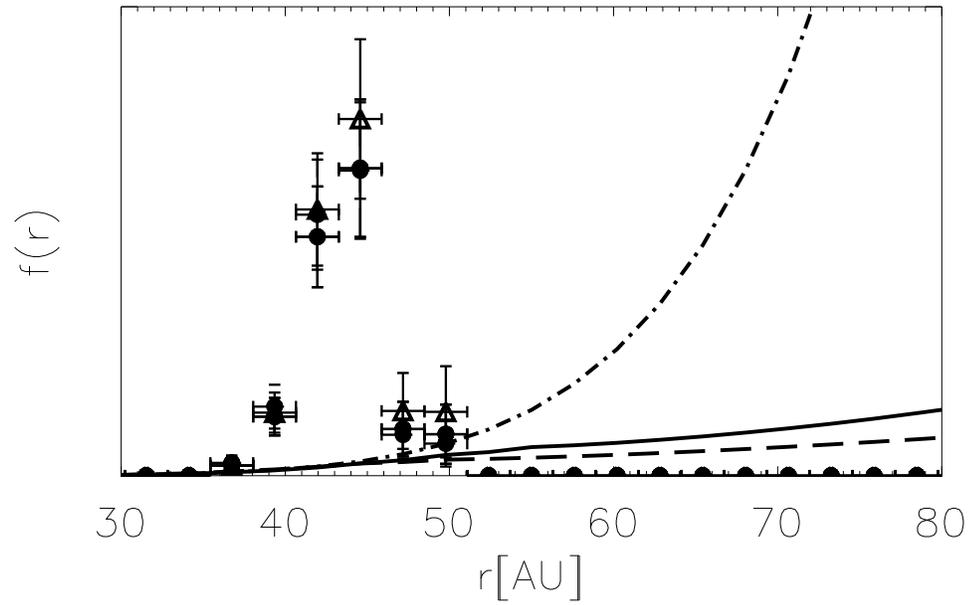}
\caption{\label{fig:distancedist} Shape of the debiased distance
  distribution of TNOs assuming a constant albedo=$4\%$. The triangles
  assume a size distribution with a power law of index $q$=4. The dark
  points assume the DPL size distributions. The bias corrections
  $\beta$ are overplot as a dot-dashed line for a single power law
  with exponent $q=4$, a solid line for a broken power law based on
  our survey and as a dashed line for parameters based on all surveys
  combined.  }
\end{figure}

Given the size distribution that corresponds to our best fit
luminosity function we are able to calculate the distance bias
in our sample and obtain the real distance distribution. We follow the
approach of \citet{Trujillo.2001}. The true and observed distributions
are related by $f(r)dr \propto \beta(r) f_o(r) dr$, where
$\beta(r)^{-1}= \int_{r_0}^{r_1} n(D) dD$ is the bias factor and
$n(D)$ is the TNO size distribution. This is done for 10 magnitude
bins between 22nd and 26th magnitude and independent estimates of the
bias function are obtained. We used the average to test the effect of
the DPL size distribution to the distance distribution of objects, as
shown in $\Fig{fig:distancedist}$. We see an abrupt drop in the
abundance of objects at $r$$\sim$47$\au$, regardless of the size
distribution considered, as has been described by others
\citep{Trujillo.2001, Petit.2006}. However, a DPL size distribution
gives a much tighter constraint on the existence of a distant
population. This is due to its much shallower size distribution for
small bodies as can be seen in the bias correction for the DPL for our
survey and the one for all surveys.

Given the fact that we detect no objects farther than 50$\au$ we can
constrain the surface density $\Sigma$ of a different population
located outside 50$\au$. At 95\% confidence level, the detection of no
objects is consistent with an expectation of 3 detected objects. We
calculated this for the observed population $N_{exp}=\Omega
\int_0^\infty \eta(x) \Sigma(x) dx$, where $\eta$ is the detection
efficiency of our survey. We assume for simplicity that the size
distribution of the distant population is the same as that we have
measured for the objects in our survey. We will also assume that each
object in the population is shifted to larger heliocentric distances
by the same factor. It is useful to define the limit on a distant
population at distance $d$ as the maximum fraction of the observed
population's surface density that a population can have to be
consistent with no detections. We denote this fraction as $g(d)$,
following the notation in \citet{Bernstein.2004}. For 60\au we find
$g=0.08$, compared to $g=0.17$ found by \citet{Bernstein.2004}. Our
survey rejects another population with the same mass closer than
$110\au$. Thus, we place a tight limit on the existance of a distant
population. We support the conclusion of \citet{Bernstein.2004} that
if such a population exists, it is either substantially less massive
than the observed classical Kuiper belt or it is comprised of small
bodies that are beyond our detection threshold.

\begin{figure}[ht]
\epsscale{1.0} \plotone{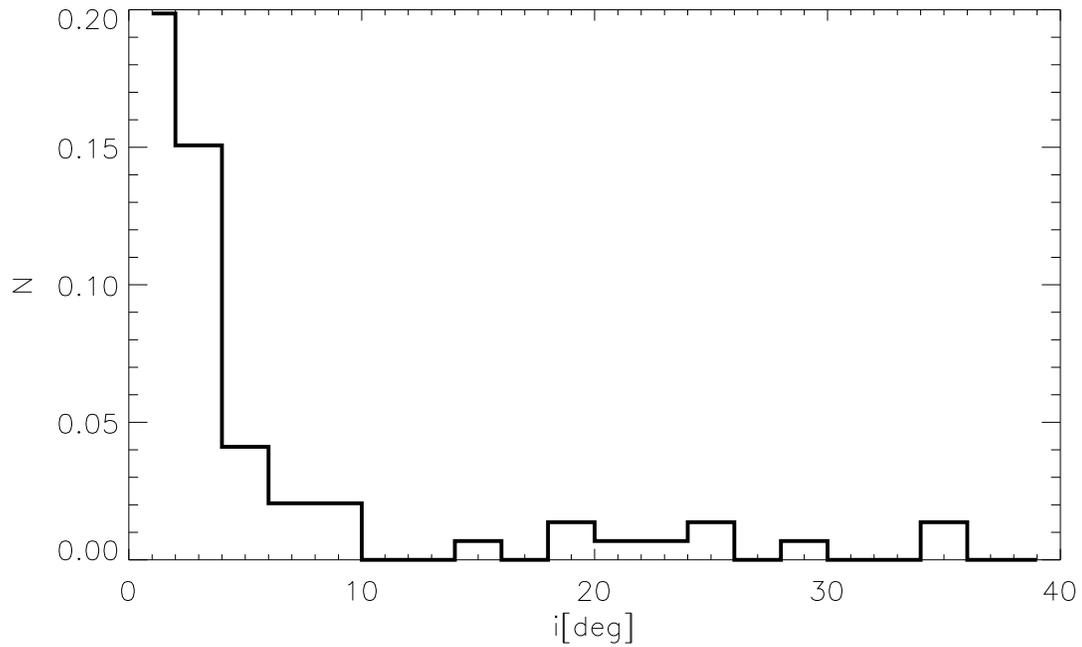}
\caption{\label{fig:inclination} Inclination probability distribution
  of TNOs in our survey. }
\end{figure}

Using the inclination information in Table \ref{tab:fits} we can show
the inclination distribution for the objects in our
survey. The results are shown in Figure \ref{fig:inclination}. This is
very similar to the results in \citet{Brown.2001}.

\subsection{Mass}\label{mass}
We use the results of our MCMC analysis to estimate the total mass of
TNOs to which our survey is sensitive.  At each step in the MCMC runs,
we compute the mass that corresponds to the DPL parameters (again,
assuming a heliocentric distance of 42$\au$ and a geometric albedo of
0.04). We follow the parametrization used in \citet{Bernstein.2004}:
\begin{eqnarray}\label{eq:mass}
  M_{tot} = M_{23} \Omega \int \sigma(R)
  10^{-0.6(R-23)}dR~f^{-1} \nonumber \\ \times
  \left[\frac{\rho}{1000~{\rm kg~m^{-3}}}\right]
  \left[\frac{d}{42\au}\right]^6
  \left[\frac{p}{0.04}\right]^{-3/2} 
\end{eqnarray}
where $M_{23}=6.3\times 10^{18}$kg = $1.055\times10^{-6} M_{\oplus}$
and $f$ is the fraction of objects from the given population that are
located within $\Omega$.

We consider the complete TNO population and the Classical and Excited
sub-samples. The DPL size distribution allows us to compute the value
of the integral in Equation \ref{eq:mass}, however the total mass of a
given population depends heavily on the mean values of the assumed
physical parameters. The mass probability distribution is calculated
assuming all other parameters are fixed. The uncertainties on the rest
of the parameters (density, albedo, distance and fraction in the
surveyed area) can be accounted for independently. We considered an
effective area of $\pm3\deg$ from the ecliptic, giving
$\Omega$=21,600$\sqdeg$ and that all objects in each population are
located within that area ($f$=1). We have also assumed mean albedo
$p=0.04$, distance $d=42\au$ and density $\rho=1000{~\rm kg~m^{-3}}$.

\begin{figure}[ht]
\epsscale{0.6} \plotone{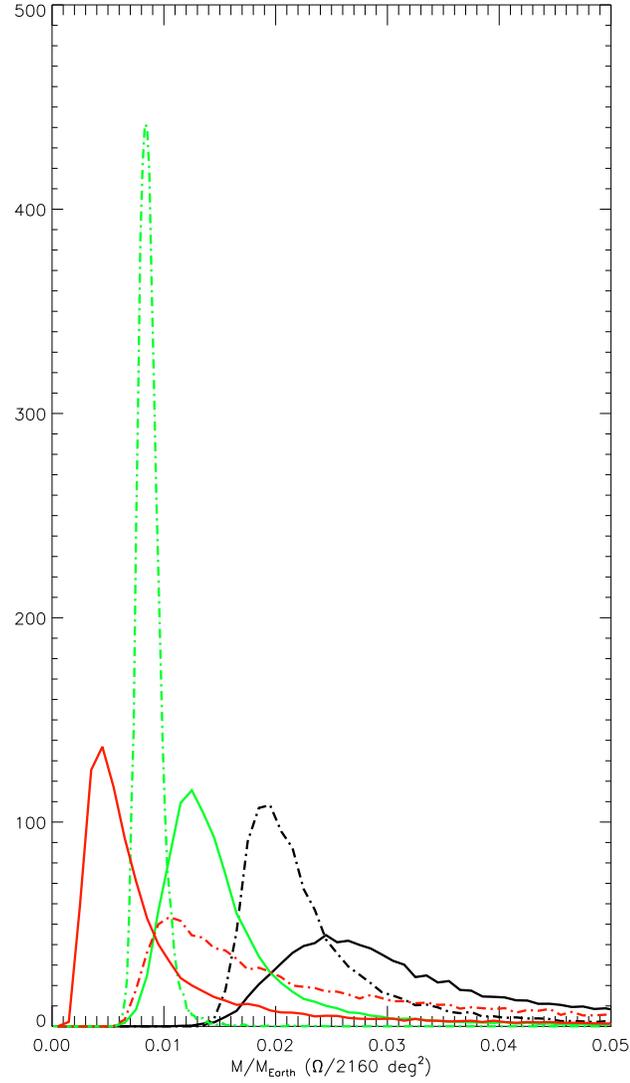}
\caption{\label{fig:massdist} Mass distribution for our population of
  TNOs. We are extending our result over a solid angle of $360 \times
  6 \sqdeg$, assuming the fraction of the population that is within
  this solid angle $f$ is 1. The solid lines represent the results
  from our survey alone while all surveys in Table \ref{tab:surv} are
  shown as dot-dashed lines. The black lines correspond to the whole
  TNO sample; green and red are used for the Classical and Excited
  sub-samples, respectively. }
\end{figure}

In Figure \ref{fig:massdist} the mass distribution is plotted for our
survey alone (solid lines) and for the combination of all the surveys
listed in Table~\ref{tab:surv} (dot-dashed lines). In black we show
the entire TNO sample. The most probable mass in TNOs for the
combination of all surveys is $M_{tno}=0.020_{-0.003}^{+0.004}
M_{\oplus}$ while for our survey alone we obtain
$M_{tno}=0.025_{-0.007}^{+0.016} M_{\oplus}$. These are consistent
with each other and with the previous estimate by
\citet{Bernstein.2004}. This is not surprising since most of the mass
is present in TNOs with sizes comparable to the size at which the
distribution breaks. The slight overabundace of TNOs in our survey
with respect to other surveys yields a higher mass for the TNO
population. It is important to note that in equation \ref{eq:mass} the
total mass diverges if either $\alpha_1<0.6$ or $\alpha_2>0.6$. We
also see in Figure \ref{fig:massdist} that for the results of our
survey alone there is a long tail to higher masses. This is due to the
poor constraint on the bright end of the TNO luminosity function given
the limited areal coverage of our survey ($2.83\sqdeg$).  However, the
combination of all surveys yields a better constraint, and we obtain
convergent masses for all steps in our MCMC run.

When we consider the Classical and Excited populations separately the
mass distributions change. In Figure \ref{fig:massdist} we show the
mass in Classical objects in green and that in Excited objects in
red. Using all the surveys the mass in classical objects is very well
constrained to be $M_{cla}=0.008\pm0.001 M_{\oplus}$.  Based on our
survey alone, we find $M_{cla}=0.013\pm0.003 M_{\oplus}$. The
overabundance of Classical objects in our survey is responsible for
that seen in the entire TNO population.

The mass in Excited objects using all surveys is $M_{exc} =
0.010_{-0.003}^{+0.021} M_{\oplus}$, larger than that found for the
Classical TNOs and is also less well constrained, with a long tail to
higher masses. This reflects the relatively poor constraint on the
size distribution of Excited objects, where the limits are set by what
values for the exponent of the power law size distribution are
considered to be physically plausible. With only 18 Excited objects in
our survey we have a very poor constraint on the individual DPL
parameters. However, the mass is well constrained.  We find $M_{exc} =
0.005_{-0.003}^{+0.004} M_{\oplus}$, less than the mass in Classical
TNOs. This is due to the relative under-abundance of Excited objects
in our survey.  This can be explained by the fact the survey was
conducted in the direction of Uranus, separated about 18.5$\deg$ from
Neptune, where we expect Plutinos to be near apocenter and hence faint
and under-represented.

\section{Conclusions}\label{sec:conc}
We have presented a TNO survey that is both deep ($R_{50}$=25.6) and
broad ($\sim$2.8$\sqdeg$), finding 82 TNOs. The survey is very
well characterized and simple, reaching its limiting magnitude in
single exposures.

We have studied the luminosity function of the TNOs in our survey. We
found a significant deviation from a single power law behavior in the
cumulative function at $R$$\sim$25. We have shown that our data are
consistent with a single power law, and with many other shallower
surveys, if we consider only objects brighter than $R$=24.5. We have
also demonstrated that the apparent deviation from a single power law
is not an artifact of our detection efficiency.

Whether our data support a break in the luminosity function is a
matter of statistical analysis. We compared two models, one where the
distribution increases exponentially with a single power law and one
where there are two different slopes in the sampled magnitude region,
and compute the total probability of each model with Bayesian
statistics \citep{Gregory.2005} (See details in Section
\ref{sec:stat}). The ratio of the total likelihood for a double power
law and a single power law model is $\sim$26. This can be interpreted
as the DPL model being 26 times more probable than the SPL given our
data set.

We conclude that our survey provides significant evidence for a break
in the TNO luminosity function. This is the first survey that is able
to make such a claim without relying upon the results of other
surveys. Our result is easy to interpret since we do not have to make
assumptions about the distribution of objects in different parts of
the sky. Nonetheless, the comparison with other surveys is fundamental
since there are published searches that sample the same magnitude
region.  We have considered most of the published data up to July 2007
regarding surveys of the trans-neptunian space in the same spirit of
\citet{Bernstein.2004}. Again, our double power law model accurately
describes the cumulative number density for all surveys combined.

Only two ground based surveys are as deep as the present, and they
have not seen a significant deviation from a single power law. The
survey of \citet{Gladman.2001} covered much less area and,
consequently, discovered many fewer TNOs in this magnitude range (17
objects for the entire survey).  Given the small numbers, our results
are not inconsistent with those of \citet{Gladman.2001}.
\citet{Fraser.2008} report the combined results of surveys taken at
different ecliptic latitudes and longitudes.  They fit for a single
power law but account for variations in the sky surface density, that
may be due to surveying at different ecliptic longitudes and
latitudes, by allowing an offset in the luminosity function zeropoint
for each survey.  This substantially increases the number of free
parameters and, we believe, allows deviations from a single power law
within individual surveys to be obscured when the results of several
surveys are combined.  We believe that this explains the difference
between the present results and those of \citet{Fraser.2008}.

We make the assumption that all objects are located at the same
distance, so the luminosity function can be translated into a size
distribution. For every object with a reliable distance estimate a
nominal size can be computed (we assume an albedo of $p$=0.04). The
size distribution of our survey was compared with the single distance
approximation and we showed they agree. We then interpret the DPL size
distribution.

The break in the size distribution reflects the size at which
collisional processes take over gravitational ones. This is, the
largest object that is expected to be disrupted in a collision in the
age of the solar system. The best DPL model for our survey features a
break at $D$=$130~(p/0.04)^{-0.5}$ km bodies while for all surveys it
is at $D$=$100~(p/0.04)^{-0.5}$ km. Current models expect the break to
occur at smaller sizes, $D\leq50\km$ for \citet{Pan.2005} and
$D\leq100\km$ for \citet{Kenyon.2004}.  We consider these models to be
consistent with our result given the assumptions on poorly constrained
quantities like the albedos on the observational side as well as
initial conditions in the theory are not well constrained. The effect
of a distribution of albedos and a possible correlation with object
size and heliocentric distances should be studied.

The inclination distribution for our survey is consistent with what is
expected from previous results \citep{Brown.2001}. However, we do not
have enough objects to do a detailed study of the distribution. We do,
however, separate our population in classical (``cold'') and excited
(``hot'') objects. We study the size distribution of these samples and
find them to show differences as done previously by
\citet{Bernstein.2004}.

We calculate the probability distribution for the total mass in TNOs,
Classical and Excited objects that are consistent with our
observations and all considered surveys. For all surveys combined we
find $M_{tno}=0.020_{-0.003}^{+0.004} M_{\oplus}$. It is interesting
to note that for the classical population the mass is very well
constrained to be $M_{cla}=0.008\pm0.001 M_{\oplus}$ while the excited
population gives a larger and poorly constrained mass of $M_{exc} =
0.010_{-0.003}^{+0.021} M_{\oplus}$. This provides evidence for a
difference between the ``hot'' and ``cold'' populations. Our survey
gives a consistent but slightly higher answer for classical objects
that we believe is due to the local overabundance of objects in our
survey. We only have 18 excited objects in our sample, too few to
constrain the parameters of the luminosity function, but enough to
show there is an under-abundance of excited objects in our
survey. This is explained by the direction of our fields, close to
where most of the Plutinos come to apocenter.

Given the size distribution we calculate a distance bias correction
\citep{Trujillo.2001a}. We then obtain the real distance distribution
of objects, assuming we are just as likely to find faint objects that
are close as those that are far. Our survey is very well suited to
detecting objects that show slow parallactic movement (distant); our
detection efficiency is essentially independent of rate for rates
larger than $0.9\aph$ (distances closer than $150\au$). According to
\citet{Dones.1997}, \citet{Jewitt.1998}, and \citet{Trujillo.2001a}
the fraction $h$ of objects found outside $48\au$ should be about 40\%
for a population with a smooth brigthness distribution that extends
beyond $50\au$. In our sample there are 73 TNOs with reliable distance
estimates, of which 71 are located between $30\au$ and $47\au$, and
only two at $\sim$$50\au$, accounting for $h$=3\%. Once we take into
account the biases associated with distance these numbers indicate an
abrupt drop in the radial density of the Kuiper Belt. If we also
consider the size distribution break found in our sample we also rule
out the existence of a far population of TNOs near the plane of the
ecliptic. We have found more evidence for an edge of the Classical
belt population at around $47\au$ and placed a constraint on the
surface density of objects for an unseen population at $60\au$ of 8\%
that of the observed Classical Belt. We also set a minimum distance
for a ``belt-like'' population with the same mass as that of the
Classical Belt of $110\au$.

Deeper surveys will help better constrain where the break in the
luminosity function occurs and complete the picture of the
trans-neptunian space. The size distribution would be better
determined if these surveys are also careful in obtaining followup
observations to measure accurate distances for faint objects.

\acknowledgments 
We are grateful to Charles Alcock, Scott Kenyon, and David Latham for
their comments and suggestions.  We thank the anonymous referee for
a very helpful and informative review.  We thank Matthew R. George for
helpful conversations and for his assistance in modifying the
\citet{Bernstein.2000} {\it Orbfit} routines. This work was supported
in part by NASA grant NNG04GK64G, issued by the NASA Planetary
Astronomy Program.


\clearpage
{\LongTables
\begin{landscape}

\begin{deluxetable}{lccccccccccc}
\tabletypesize{\scriptsize}
\tablecaption{\sc  Fit Parameters\tablenotemark{*}}
\tablewidth{0pt}
\tablehead{
  \colhead{$Name$} &
  \colhead{${\rm MJD}$} &
  \colhead{${\rm RA}$} &
  \colhead{${\rm Dec}$} &
  \colhead{$R_{mag}$} &
  \colhead{$\Delta_{Rmag}$} &
  \colhead{$d{\rm RA}/dt$} &
  \colhead{$d{\rm Dec}/dt$} &
  \colhead{$d_{par}$} &
  \colhead{$d_{bari}$} &
  \colhead{$i$} &
  \colhead{$\Delta_{\alpha}(Uranus)$} \\
  \colhead{} &
  \colhead{} &
  \colhead{} &
  \colhead{} &
  \colhead{} &
  \colhead{} &
  \colhead{$[''/hr]$} &
  \colhead{$[''/hr]$} &
  \colhead{$[AU]$} &
  \colhead{$[AU]$} &
  \colhead{$[deg]$} &
  \colhead{$[']$}
}
\startdata
sukbo88 & $52880.456823$ & $22:09:09.42$ & $-12:46:24.35$ & $23.57_{-0.05}^{+0.05}$ & $0.07$ & $-2.81$ & $-1.04$ & $43.0$ & $42.9 \pm 2.5$ & $2.2 \pm 1.1$ & $60$ \\ 
sukbo57 & $52880.387131$ & $22:09:15.90$ & $-11:55:25.85$ & $24.97_{-0.15}^{+0.13}$ & $0.14$ & $-2.78$ & $-1.02$ & $43.4$ & $43.2 \pm 2.5$ & $0.7 \pm 0.8$ & $38$ \\ 
sukbo17 & $52880.467440$ & $22:09:24.36$ & $-11:37:45.60$ & $25.03_{-0.16}^{+0.14}$ & $0.14$ & $-2.67$ & $-1.01$ & $45.2$ & $44.9 \pm 2.5$ & $2.7 \pm 1.6$ & $42$ \\ 
sukbo23 & $52880.467440$ & $22:09:26.17$ & $-11:16:08.70$ & $24.25_{-0.11}^{+0.10}$ & $0.10$ & $-2.68$ & $-0.90$ & $46.0$ & $45.5 \pm 2.6$ & $9.4 \pm 3.6$ & $56$ \\ 
sukbo59 & $52880.387131$ & $22:09:33.47$ & $-11:52:17.45$ & $25.47_{-0.19}^{+0.16}$ & $0.17$ & $-2.82$ & $-1.04$ & $42.9$ & $42.6 \pm 2.5$ & $1.0 \pm 1.1$ & $34$ \\ 
sukbo52 & $52880.387131$ & $22:09:36.21$ & $-12:06:01.14$ & $23.75_{-0.04}^{+0.04}$ & $0.08$ & $-2.58$ & $-0.96$ & $47.0$ & $46.9 \pm 2.5$ & $1.6 \pm 1.3$ & $33$ \\ 
sukbo90 & $52880.456823$ & $22:09:38.29$ & $-12:39:41.88$ & $25.60_{-0.23}^{+0.19}$ & $0.17$ & $-2.95$ & $-1.09$ & $40.8$ & $40.6 \pm 2.5$ & $1.7 \pm 0.8$ & $51$ \\ 
sukbo24 & $52880.467440$ & $22:09:40.30$ & $-11:12:11.25$ & $24.54_{-0.13}^{+0.12}$ & $0.12$ & $-2.70$ & $-1.00$ & $44.8$ & $44.5 \pm 2.5$ & $0.5 \pm 1.2$ & $58$ \\ 
sukbo51 & $52880.387131$ & $22:09:45.05$ & $-12:08:45.55$ & $24.37_{-0.08}^{+0.08}$ & $0.11$ & $-3.05$ & $-1.11$ & $39.5$ & $39.2 \pm 2.4$ & $1.0 \pm 0.7$ & $31$ \\ 
sukbo50 & $52880.387131$ & $22:09:47.61$ & $-12:10:06.91$ & $25.13_{-0.17}^{+0.15}$ & $0.15$ & $-2.73$ & $-0.98$ & $44.6$ & $44.3 \pm 2.5$ & $2.4 \pm 1.4$ & $31$ \\ 
sukbo54 & $52880.387131$ & $22:09:48.36$ & $-12:05:40.42$ & $25.59_{-0.25}^{+0.20}$ & $0.17$ & $-2.68$ & $-0.99$ & $45.2$ & $45.0 \pm 2.5$ & $1.5 \pm 1.2$ & $30$ \\ 
sukbo22 & $52880.467440$ & $22:09:51.25$ & $-11:21:10.92$ & $24.74_{-0.15}^{+0.13}$ & $0.13$ & $-2.36$ & $-1.02$ & $50.8$ & $50.8 \pm 2.8$ & $18.4 \pm 7.0$ & $49$ \\ 
sukbo55 & $52880.387131$ & $22:09:51.26$ & $-12:03:05.07$ & $24.63_{-0.14}^{+0.12}$ & $0.12$ & $-2.69$ & $-1.03$ & $44.7$ & $44.6 \pm 2.5$ & $4.2 \pm 1.9$ & $29$ \\ 
sukbo48 & $52880.387131$ & $22:09:53.94$ & $-12:16:51.14$ & $25.52_{-0.20}^{+0.17}$ & $0.17$ & $-3.13$ & $-1.54$ & $37.6$ & $37.9 \pm 3.3$ & $34.6 \pm 16.4$ & $32$ \\ 
sukbo60$^a$ & $52880.387131$ & $22:09:55.42$ & $-11:50:18.49$ & $25.77_{-0.23}^{+0.19}$ & $0.19$ & $-2.82$ & $-1.03$ & $42.9$ & $46.4 \pm 12.0$ & $^b$ & $29$ \\ 
sukbo58$^a$ & $52880.387131$ & $22:10:01.46$ & $-11:52:42.62$ & $25.68_{-0.25}^{+0.20}$ & $0.18$ & $-2.85$ & $-0.86$ & $44.3$ & $47.5 \pm 11.1$ & $^b$ & $27$ \\ 
sukbo91 & $52880.456823$ & $22:10:26.15$ & $-12:35:13.32$ & $24.36_{-0.08}^{+0.07}$ & $0.11$ & $-2.76$ & $-1.06$ & $43.5$ & $43.4 \pm 2.5$ & $5.0 \pm 2.0$ & $40$ \\ 
sukbo21 & $52880.467440$ & $22:10:27.48$ & $-11:26:00.88$ & $23.72_{-0.06}^{+0.06}$ & $0.07$ & $-2.79$ & $-1.06$ & $43.1$ & $42.9 \pm 2.5$ & $3.2 \pm 1.6$ & $40$ \\ 
sukbo16 & $52880.467440$ & $22:10:30.60$ & $-11:41:46.06$ & $23.31_{-0.05}^{+0.04}$ & $0.06$ & $-2.82$ & $-1.00$ & $43.2$ & $42.8 \pm 2.5$ & $3.9 \pm 1.8$ & $27$ \\ 
sukbo53 & $52880.387131$ & $22:10:31.56$ & $-12:06:20.06$ & $24.52_{-0.09}^{+0.08}$ & $0.11$ & $-2.57$ & $-0.95$ & $47.2$ & $47.0 \pm 2.5$ & $0.8 \pm 0.1$ & $20$ \\ 
sukbo56 & $52880.387131$ & $22:10:32.98$ & $-12:02:16.43$ & $23.95_{-0.05}^{+0.05}$ & $0.09$ & $-2.67$ & $-1.01$ & $45.2$ & $45.2 \pm 2.5$ & $3.9 \pm 1.8$ & $18$ \\ 
sukbo93 & $52880.456823$ & $22:10:36.60$ & $-12:18:23.83$ & $23.97_{-0.09}^{+0.08}$ & $0.09$ & $-2.69$ & $-1.00$ & $44.9$ & $44.8 \pm 2.5$ & $1.5 \pm 1.1$ & $25$ \\ 
sukbo92 & $52880.456823$ & $22:10:39.91$ & $-12:26:38.52$ & $25.38_{-0.27}^{+0.21}$ & $0.16$ & $-2.87$ & $-1.07$ & $42.0$ & $41.8 \pm 2.5$ & $2.7 \pm 1.3$ & $31$ \\ 
sukbo94 & $52880.456823$ & $22:10:42.50$ & $-12:18:33.94$ & $23.85_{-0.07}^{+0.07}$ & $0.08$ & $-2.79$ & $-1.01$ & $43.5$ & $43.3 \pm 2.5$ & $1.7 \pm 1.1$ & $24$ \\ 
sukbo45 & $52880.348317$ & $22:10:51.30$ & $-12:36:53.32$ & $24.15_{-0.07}^{+0.06}$ & $0.10$ & $-2.92$ & $-1.09$ & $41.2$ & $41.1 \pm 2.5$ & $2.3 \pm 1.0$ & $39$ \\ 
sukbo49$^a$ & $52880.387131$ & $22:10:52.67$ & $-12:13:42.84$ & $25.41_{-0.20}^{+0.17}$ & $0.16$ & $-2.89$ & $-0.95$ & $42.6$ & $46.8 \pm 12.0$ & $^b$ & $19$ \\ 
sukbo0 & $52880.337272$ & $22:10:52.89$ & $-12:13:41.26$ & $25.23_{-0.18}^{+0.15}$ & $0.15$ & $-2.84$ & $-1.03$ & $42.7$ & $42.5 \pm 2.5$ & $1.3 \pm 0.8$ & $19$ \\ 
sukbo61$^a$ & $52880.387131$ & $22:10:53.87$ & $-11:45:27.66$ & $23.13_{-0.03}^{+0.03}$ & $0.05$ & $-2.88$ & $-1.08$ & $41.7$ & $43.1 \pm 9.8$ & $^b$ & $20$ \\ 
sukbo31 & $52880.342793$ & $22:10:54.08$ & $-11:45:26.57$ & $23.19_{-0.03}^{+0.03}$ & $0.05$ & $-2.84$ & $-1.06$ & $42.4$ & $42.2 \pm 2.5$ & $1.2 \pm 1.1$ & $20$ \\ 
sukbo2 & $52880.337272$ & $22:10:54.93$ & $-12:12:09.59$ & $25.64_{-0.24}^{+0.19}$ & $0.18$ & $-2.72$ & $-1.05$ & $44.2$ & $44.2 \pm 2.5$ & $6.1 \pm 2.4$ & $17$ \\ 
sukbo44 & $52880.348317$ & $22:10:57.80$ & $-12:42:51.61$ & $25.25_{-0.16}^{+0.14}$ & $0.15$ & $-3.11$ & $-1.25$ & $38.2$ & $38.1 \pm 2.5$ & $9.8 \pm 3.7$ & $44$ \\ 
sukbo34 & $52880.342793$ & $22:11:03.64$ & $-11:31:33.24$ & $24.17_{-0.07}^{+0.07}$ & $0.10$ & $-2.77$ & $-1.03$ & $43.5$ & $43.3 \pm 2.5$ & $0.9 \pm 1.1$ & $31$ \\ 
sukbo27 & $52880.472734$ & $22:11:04.46$ & $-13:09:42.98$ & $25.16_{-0.16}^{+0.14}$ & $0.15$ & $-2.94$ & $-1.09$ & $41.0$ & $41.0 \pm 2.5$ & $2.7 \pm 0.9$ & $70$ \\ 
sukbo73 & $52880.397724$ & $22:11:06.70$ & $-10:54:01.82$ & $24.96_{-0.15}^{+0.13}$ & $0.14$ & $-2.65$ & $-1.01$ & $45.5$ & $45.3 \pm 2.5$ & $2.3 \pm 1.5$ & $67$ \\ 
sukbo46 & $52880.348317$ & $22:11:09.54$ & $-12:35:08.71$ & $24.58_{-0.10}^{+0.10}$ & $0.12$ & $-2.59$ & $-0.99$ & $46.6$ & $46.7 \pm 2.5$ & $5.5 \pm 2.2$ & $36$ \\ 
sukbo25 & $52880.472734$ & $22:11:15.88$ & $-13:17:16.77$ & $24.95_{-0.21}^{+0.17}$ & $0.14$ & $-3.13$ & $-1.24$ & $37.9$ & $38.0 \pm 2.5$ & $8.6 \pm 3.1$ & $77$ \\ 
sukbo39 & $52880.348317$ & $22:11:17.04$ & $-12:51:56.40$ & $25.37_{-0.20}^{+0.17}$ & $0.16$ & $-2.79$ & $-1.04$ & $43.2$ & $43.2 \pm 2.5$ & $2.3 \pm 1.0$ & $52$ \\ 
sukbo8 & $52880.337272$ & $22:11:20.06$ & $-12:03:12.63$ & $25.13_{-0.15}^{+0.13}$ & $0.15$ & $-2.80$ & $-1.05$ & $43.1$ & $42.9 \pm 2.5$ & $1.7 \pm 1.1$ & $7$ \\ 
sukbo6 & $52880.337272$ & $22:11:23.29$ & $-12:05:17.03$ & $25.80_{-0.27}^{+0.21}$ & $0.19$ & $-2.83$ & $-1.02$ & $42.9$ & $42.7 \pm 2.5$ & $1.7 \pm 1.1$ & $7$ \\ 
sukbo33 & $52880.342793$ & $22:11:24.19$ & $-11:37:12.39$ & $24.19_{-0.07}^{+0.06}$ & $0.10$ & $-2.96$ & $-1.06$ & $40.8$ & $40.6 \pm 2.5$ & $2.8 \pm 1.4$ & $24$ \\ 
sukbo43 & $52880.348317$ & $22:11:24.34$ & $-12:48:33.44$ & $25.18_{-0.20}^{+0.16}$ & $0.15$ & $-2.84$ & $-1.29$ & $41.7$ & $42.1 \pm 2.9$ & $25.5 \pm 10.3$ & $48$ \\ 
sukbo3 & $52880.337272$ & $22:11:26.81$ & $-12:11:39.44$ & $25.60_{-0.22}^{+0.18}$ & $0.18$ & $-2.75$ & $-1.04$ & $43.8$ & $43.7 \pm 2.5$ & $3.2 \pm 1.5$ & $12$ \\ 
sukbo42 & $52880.348317$ & $22:11:37.53$ & $-12:49:36.33$ & $24.55_{-0.11}^{+0.10}$ & $0.12$ & $-2.87$ & $-1.03$ & $42.3$ & $42.1 \pm 2.5$ & $2.6 \pm 1.1$ & $49$ \\ 
sukbo77 & $52880.397724$ & $22:11:47.57$ & $-10:51:02.90$ & $23.34_{-0.03}^{+0.03}$ & $0.06$ & $-2.83$ & $-1.06$ & $42.5$ & $42.2 \pm 2.5$ & $1.0 \pm 1.1$ & $69$ \\ 
sukbo32 & $52880.342793$ & $22:11:47.88$ & $-11:38:28.77$ & $24.70_{-0.10}^{+0.09}$ & $0.12$ & $-2.79$ & $-1.07$ & $43.1$ & $43.0 \pm 2.5$ & $3.7 \pm 1.7$ & $22$ \\ 
sukbo1 & $52880.337272$ & $22:11:49.55$ & $-12:12:54.47$ & $25.62_{-0.22}^{+0.18}$ & $0.18$ & $-2.78$ & $-1.05$ & $43.3$ & $43.3 \pm 2.5$ & $3.1 \pm 1.4$ & $13$ \\ 
sukbo5 & $52880.337272$ & $22:11:51.99$ & $-12:07:17.95$ & $24.27_{-0.07}^{+0.07}$ & $0.10$ & $-2.89$ & $-1.07$ & $41.6$ & $41.5 \pm 2.5$ & $1.2 \pm 0.8$ & $7$ \\ 
sukbo37$^a$ & $52880.342793$ & $22:11:53.44$ & $-11:26:56.13$ & $25.65_{-0.26}^{+0.21}$ & $0.18$ & $-2.38$ & $-0.59$ & $60.1$ & $^b$ & $^b$ & $34$ \\ 
sukbo4 & $52880.337272$ & $22:11:53.69$ & $-12:10:54.32$ & $24.46_{-0.08}^{+0.08}$ & $0.11$ & $-3.19$ & $-1.54$ & $36.7$ & $36.9 \pm 3.0$ & $29.8 \pm 13.2$ & $11$ \\ 
sukbo13 & $52880.337272$ & $22:12:00.29$ & $-11:59:26.11$ & $25.37_{-0.21}^{+0.17}$ & $0.16$ & $-2.81$ & $-1.38$ & $42.1$ & $42.4 \pm 3.4$ & $35.2 \pm 16.2$ & $4$ \\ 
sukbo38 & $52880.342793$ & $22:12:08.82$ & $-11:16:55.93$ & $25.29_{-0.18}^{+0.15}$ & $0.16$ & $-2.51$ & $-0.76$ & $50.6$ & $50.0 \pm 2.8$ & $21.0 \pm 8.0$ & $44$ \\ 
sukbo28 & $52880.472734$ & $22:12:21.36$ & $-13:01:26.73$ & $23.39_{-0.05}^{+0.05}$ & $0.06$ & $-2.77$ & $-1.03$ & $43.6$ & $43.6 \pm 2.5$ & $2.6 \pm 1.0$ & $62$ \\ 
sukbo78 & $52880.397724$ & $22:12:21.47$ & $-10:42:48.31$ & $22.64_{-0.02}^{+0.02}$ & $0.03$ & $-2.78$ & $-1.10$ & $43.0$ & $42.9 \pm 2.5$ & $6.6 \pm 2.7$ & $78$ \\ 
sukbo74 & $52880.397724$ & $22:12:27.15$ & $-10:52:56.00$ & $22.99_{-0.02}^{+0.02}$ & $0.04$ & $-2.72$ & $-1.00$ & $44.5$ & $44.3 \pm 2.5$ & $1.7 \pm 1.3$ & $68$ \\ 
sukbo76 & $52880.397724$ & $22:12:27.37$ & $-10:51:42.55$ & $25.49_{-0.21}^{+0.17}$ & $0.17$ & $-2.86$ & $-1.10$ & $41.9$ & $41.6 \pm 2.5$ & $3.1 \pm 1.6$ & $69$ \\ 
sukbo26 & $52880.472734$ & $22:12:27.91$ & $-13:17:09.62$ & $25.17_{-0.20}^{+0.17}$ & $0.15$ & $-2.86$ & $-1.12$ & $41.8$ & $41.8 \pm 2.5$ & $7.8 \pm 2.8$ & $77$ \\ 
sukbo75 & $52880.397724$ & $22:12:28.22$ & $-10:51:23.92$ & $25.59_{-0.20}^{+0.17}$ & $0.17$ & $-2.84$ & $-0.84$ & $44.7$ & $44.1 \pm 2.8$ & $23.3 \pm 9.3$ & $70$ \\ 
sukbo41 & $52880.348317$ & $22:12:28.24$ & $-12:50:19.12$ & $24.92_{-0.15}^{+0.13}$ & $0.14$ & $-2.95$ & $-1.14$ & $40.5$ & $40.5 \pm 2.5$ & $5.4 \pm 2.0$ & $51$ \\ 
sukbo29 & $52880.472734$ & $22:12:28.90$ & $-13:00:29.66$ & $23.86_{-0.07}^{+0.06}$ & $0.08$ & $-2.91$ & $-1.07$ & $41.4$ & $41.4 \pm 2.5$ & $1.9 \pm 0.5$ & $61$ \\ 
sukbo35 & $52880.342793$ & $22:12:32.78$ & $-11:31:13.39$ & $25.58_{-0.20}^{+0.17}$ & $0.17$ & $-2.72$ & $-1.25$ & $43.5$ & $43.6 \pm 2.9$ & $25.3 \pm 10.3$ & $31$ \\ 
sukbo99 & $52880.462148$ & $22:12:40.64$ & $-12:27:42.77$ & $24.28_{-0.09}^{+0.09}$ & $0.10$ & $-2.76$ & $-1.01$ & $43.9$ & $43.8 \pm 2.5$ & $1.3 \pm 0.1$ & $31$ \\ 
sukbo47$^a$ & $52880.348317$ & $22:12:41.16$ & $-12:27:39.37$ & $23.89_{-0.05}^{+0.05}$ & $0.08$ & $-2.81$ & $-1.07$ & $42.8$ & $^b$ & $^b$ & $31$ \\ 
sukbo81 & $52880.403236$ & $22:12:54.24$ & $-11:39:25.88$ & $25.26_{-0.18}^{+0.15}$ & $0.16$ & $-3.06$ & $-1.34$ & $38.5$ & $38.5 \pm 2.6$ & $18.1 \pm 7.1$ & $27$ \\ 
sukbo69 & $52880.392433$ & $22:13:01.10$ & $-11:52:24.31$ & $25.26_{-0.18}^{+0.15}$ & $0.16$ & $-2.98$ & $-1.10$ & $40.4$ & $40.3 \pm 2.4$ & $0.8 \pm 0.3$ & $20$ \\ 
sukbo67 & $52880.392433$ & $22:13:02.05$ & $-12:08:55.50$ & $24.66_{-0.10}^{+0.09}$ & $0.12$ & $-2.88$ & $-1.04$ & $42.1$ & $41.9 \pm 2.5$ & $2.0 \pm 1.1$ & $21$ \\ 
sukbo64 & $52880.392433$ & $22:13:07.07$ & $-12:12:25.08$ & $24.48_{-0.09}^{+0.08}$ & $0.11$ & $-2.92$ & $-1.11$ & $41.0$ & $41.0 \pm 2.5$ & $2.6 \pm 1.2$ & $24$ \\ 
sukbo85 & $52880.403236$ & $22:13:16.86$ & $-11:25:45.74$ & $24.20_{-0.07}^{+0.06}$ & $0.10$ & $-2.76$ & $-1.04$ & $43.7$ & $43.6 \pm 2.5$ & $1.9 \pm 1.3$ & $41$ \\ 
sukbo63 & $52880.392433$ & $22:13:18.02$ & $-12:13:17.31$ & $25.37_{-0.19}^{+0.16}$ & $0.16$ & $-2.81$ & $-1.05$ & $42.9$ & $42.9 \pm 2.5$ & $1.9 \pm 1.0$ & $26$ \\ 
sukbo65 & $52880.392433$ & $22:13:25.44$ & $-12:09:50.02$ & $25.52_{-0.19}^{+0.16}$ & $0.17$ & $-2.69$ & $-1.00$ & $45.0$ & $44.9 \pm 2.5$ & $1.6 \pm 1.0$ & $27$ \\ 
sukbo87$^a$ & $52880.403236$ & $22:13:26.02$ & $-11:17:06.76$ & $25.27_{-0.18}^{+0.15}$ & $0.16$ & $-2.62$ & $-0.94$ & $46.5$ & $^b$ & $^b$ & $50$ \\ 
sukbo96 & $52880.462148$ & $22:13:28.31$ & $-12:42:21.29$ & $25.40_{-0.26}^{+0.21}$ & $0.16$ & $-2.74$ & $-0.98$ & $44.4$ & $44.2 \pm 2.5$ & $3.3 \pm 1.4$ & $49$ \\ 
sukbo100 & $52880.462148$ & $22:13:39.25$ & $-12:26:53.85$ & $24.31_{-0.09}^{+0.08}$ & $0.10$ & $-2.99$ & $-1.05$ & $40.6$ & $40.4 \pm 2.5$ & $4.3 \pm 1.8$ & $39$ \\ 
sukbo62$^a$ & $52880.392433$ & $22:13:42.77$ & $-12:15:37.04$ & $25.16_{-0.14}^{+0.13}$ & $0.15$ & $-2.01$ & $-1.34$ & $60.1$ & $65.2 \pm 17.4$ & $^b$ & $33$ \\ 
sukbo72 & $52880.392433$ & $22:13:51.41$ & $-11:46:46.76$ & $25.14_{-0.17}^{+0.14}$ & $0.15$ & $-2.81$ & $-1.05$ & $42.9$ & $42.8 \pm 2.5$ & $1.0 \pm 0.7$ & $34$ \\ 
sukbo80 & $52880.403236$ & $22:13:56.31$ & $-11:40:41.22$ & $25.38_{-0.22}^{+0.18}$ & $0.16$ & $-2.70$ & $-1.02$ & $44.7$ & $44.6 \pm 2.5$ & $1.8 \pm 1.2$ & $38$ \\ 
sukbo86 & $52880.403236$ & $22:13:58.17$ & $-11:23:21.45$ & $24.94_{-0.14}^{+0.12}$ & $0.14$ & $-3.13$ & $-0.99$ & $39.5$ & $39.1 \pm 2.5$ & $14.8 \pm 5.8$ & $50$ \\ 
sukbo79 & $52880.403236$ & $22:13:59.82$ & $-11:41:10.37$ & $24.61_{-0.17}^{+0.14}$ & $0.12$ & $-3.06$ & $-1.19$ & $39.0$ & $38.9 \pm 2.4$ & $4.7 \pm 1.9$ & $39$ \\ 
sukbo95 & $52880.462148$ & $22:14:02.62$ & $-12:47:04.86$ & $24.71_{-0.16}^{+0.14}$ & $0.12$ & $-2.81$ & $-1.02$ & $43.1$ & $43.1 \pm 2.5$ & $1.7 \pm 0.3$ & $58$ \\ 
sukbo70 & $52880.392433$ & $22:14:06.42$ & $-11:48:59.37$ & $25.04_{-0.14}^{+0.12}$ & $0.14$ & $-2.76$ & $-1.04$ & $43.6$ & $43.5 \pm 2.5$ & $2.1 \pm 1.2$ & $37$ \\ 
sukbo71 & $52880.392433$ & $22:14:13.53$ & $-11:47:47.21$ & $24.25_{-0.09}^{+0.08}$ & $0.10$ & $-3.00$ & $-1.09$ & $40.2$ & $40.0 \pm 2.4$ & $1.9 \pm 1.1$ & $39$ \\ 
sukbo97 & $52880.462148$ & $22:14:13.95$ & $-12:38:36.76$ & $24.73_{-0.13}^{+0.12}$ & $0.13$ & $-2.84$ & $-1.06$ & $42.4$ & $42.4 \pm 2.5$ & $2.1 \pm 0.8$ & $53$ \\ 
sukbo66$^a$ & $52880.392433$ & $22:14:22.17$ & $-12:09:05.58$ & $25.01_{-0.15}^{+0.13}$ & $0.14$ & $-3.32$ & $-1.65$ & $35.2$ & $34.9 \pm 7.5$ & $^b$ & $40$ \\ 
\enddata
\tablenotetext{*}{ All 82 trans-neptunian objects found. The second night data
  were used when possible. The measured magnitude in the R filter with
  nominal errors is shown in $R_{mag}$. $\Delta_{Rmag}$ is a model for
  the photometric error based on the measure magnitudes of inserted,
  synthetic objects. $d{\rm RA}/dt$ and
  $d{\rm Dec}/dt$ are estimates of the measured motion of the object. The
  distance $d_{par}$ is calculated with the assumption of a circular
  orbit. $d_{bari}$ is the barycentric distance estimate and $i$ is
  the inclination estimate given by the {\it Orbfit} code
  \citep{Bernstein.2000}. $\Delta_{\alpha}(Uranus)$ is the projected
  distance to Uranus during the observations.  }
\tablenotetext{a}{ These objects were not found in the second night of observations.}
\tablenotetext{b}{ The result is unconstrained.}
\label{tab:fits}
\end{deluxetable}

\clearpage
\end{landscape}
}

\clearpage
\newpage

\begin{deluxetable}{lcccccc}
\tabletypesize{\scriptsize}
\tablecaption{\sc Surveys\tablenotemark{*}}
\tablewidth{0pt}
\tablehead{
  \colhead{$Paper$} &
  \colhead{$\Omega$} &
  \colhead{$R_{50}$} &
  \colhead{$N_{C}$\tablenotemark{a}} &
  \colhead{$N_{E}$\tablenotemark{a}} &
  \colhead{$N_{obs}$} &
  \colhead{$N_{exp}$} \\
  \colhead{} &
  \colhead{$\sqdeg$} &
  \colhead{} &
  \colhead{} &
  \colhead{} &
  \colhead{} &
  \colhead{}
}
\startdata
\cite{Chiang.1999}\tablenotemark{b} & 0.01 & 27.0 & 1 & 1 & 2 & 1 \\
\cite{Gladman.2001}\tablenotemark{c} & 0.322 & 25.9 & 7 & 8 & 15 & 15\\
\cite{Trujillo.2001}\tablenotemark{d} & 28.3 & 23.7 & 38 & 27 & 71 & 64\\
\cite{Allen.2002} & 2.30 & 25.1 & 15 & 15 & 30 & 39\\
\cite{Bernstein.2004} & 0.019 & 28.5 & 3 & 0 & 3 & 5\\
\cite{Petit.2006} N & 5.88 & 24.2 & 6 & 21 & 27 & 22\\
\cite{Petit.2006} U & 5.97 & 24.6 & 16 & 20 & 36 & 34\\
\cite{Fraser.2008} & 3.0 & 20.8 & 36 & 31 & 67 & 74\\
This survey\tablenotemark{d} & 2.83 & 25.69 & 54 & 18 & 82 & 74\\
\enddata
\tablenotetext{*}{Details of the surveys considered in
  this work. $\Omega$ is the total surveyed area. $R_{50}$ defines the
  $R$ magnitude at which the survey's detection efficiency is 50\% its
  maximum efficiency. The total number of objects discovered that had
  magnitude brighter than that at which the survey is 15\% its maximum
  efficiency is $N_{obs}$, as defined in \citet{Bernstein.2004}. The
  expected number of objects for each survey given our most likely DPL
  luminosity function model for all surveys combined (see
  Fig. \ref{fig:like2all}) is $N_{exp}$.  }
\tablenotetext{a}{Objects with inclination $i\leq 5\deg$ and at a distance
  $38\au < d < 55\au$ are considered as Classical $N_C$ and the rest as
  Excited $N_E$.}
\tablenotetext{b}{Based on Table 3 and comments in \cite{Gladman.2001}.}
\tablenotetext{c}{Based on Table 2 and comments in \cite{Bernstein.2004}.}
\tablenotetext{d}{In the Classical and Extended classification We only
  considered objects for which there was distance and inclination
  information.}
\label{tab:surv}
\end{deluxetable}

\begin{deluxetable}{llcccc}
\tabletypesize{\scriptsize}
\tablecaption{\sc DPL Parameter Estimation\tablenotemark{*}}
\tablewidth{0pt}
\tablehead{
  \colhead{$Survey$} &
  \colhead{$$} &
  \colhead{$\alpha_1$} &
  \colhead{$\alpha_2$} &
  \colhead{$\sigma_{23}$} &
  \colhead{$R_{eq}$}
}
\startdata
All surveys & TNO & $0.75_{-0.08}^{+0.12}$ & $0.23_{-0.14}^{+0.07}$ & $1.50_{-0.12}^{+0.18}$ & $24.8_{-0.9}^{+0.5}$ \\
 & Classical & $1.4_{-0.3}^{+0.1}$ & $0.32_{-0.06}^{+0.04}$ & $0.82_{-0.12}^{+0.13}$ & $23.3_{-0.3}^{+0.3}$ \\
 & Excited & $0.61_{-0.05}^{+0.07}$ & $-0.3_{-0.2}^{+0.4}$ & $0.68_{-0.08}^{+0.09}$ & $25.7_{-0.6}^{+0.7}$ \\
This survey$_\dagger$ & TNO & $0.7_{-0.1}^{+0.2}$ & $0.3_{-0.2}^{+0.2}$ & $2.0_{-0.5}^{+0.5}$ & $24.3_{-0.1}^{+0.8}$ \\
 & Classical & $1.2_{-0.4}^{+0.3}$ & $0.15_{-0.15}^{+0.20}$ & $1.5_{-0.5}^{+0.5}$ & $23.6_{-0.7}^{+0.6}$ \\
\enddata
\tablenotetext{*}{Best fit parameters and 1-$\sigma$ confidence limits
  based on MCMC simulations. All surveys are detailed in Table
  \ref{tab:surv}. }
\tablenotetext{$\dagger$}{In this survey there were only 18 excited
  objects, too few to constrain a 4-parameter model. However we could
  fit a SPL with $\alpha$=$0.62\pm0.12$ and $R_0$=$24.2\pm0.3$ to this
  population. }
\label{tab:params}
\end{deluxetable}

\clearpage
\newpage

\clearpage
\newpage

\end{document}